\documentclass[prb,twocolumn,superscriptaddress]{revtex4}
\usepackage{amsmath, amsfonts,amssymb,color,graphicx}

\begin{document} 

\title{Hot carrier extraction with plasmonic broadband absorbers}

\author{Charlene Ng}
\email{charlene.ng@csiro.au}
\affiliation{CSIRO, Manufacturing,  Private Bag 33, Clayton, VIC, 3168, Australia}
\affiliation{Melbourne Centre for Nanofabrication, Australian National Fabrication Facility,  Clayton VIC 3168, Australia}

\author{Jasper Cadusch}
\affiliation{School of Physics, The University of Melbourne, Parkville, VIC, 3010, Australia}

\author{Svetlana Dligatch}
\affiliation{CSIRO, Manufacturing, PO Box 218, Lindfield NSW 2070, Australia}

\author{Ann Roberts}
\affiliation{School of Physics, The University of Melbourne, Parkville, VIC, 3010, Australia}

\author{Timothy J. Davis}
\affiliation{School of Physics, The University of Melbourne, Parkville, VIC, 3010, Australia}
\affiliation{Melbourne Centre for Nanofabrication, Australian National Fabrication Facility,  Clayton VIC 3168, Australia}

\author{Paul Mulvaney}
\affiliation{Bio21 Institute \& School of Chemistry, The University of Melbourne, Parkville VIC 3010, Australia}

\author{Daniel E. G\'omez}
\email{daniel.gomez@csiro.au}
\affiliation{CSIRO, Manufacturing,  Private Bag 33, Clayton, VIC, 3168, Australia}
\affiliation{Melbourne Centre for Nanofabrication, Australian National Fabrication Facility,  Clayton VIC 3168, Australia}
\affiliation{School of Physics, The University of Melbourne, Parkville, VIC, 3010, Australia}

\keywords{Surface plasmons, hot--spots, surface enhancements, near fields, photocatalysis}

\maketitle

%

\textbf{
Hot charge carrier extraction from  metallic nanostructures is a very promising approach for applications in photo-catalysis, photovoltaics and photodetection. 
One  limitation is that many metallic nanostructures support a single plasmon resonance thus restricting the light-to-charge-carrier activity to a spectral band. 
Here we demonstrate that a monolayer of plasmonic nanoparticles can be assembled on a multi–stack layered configuration to achieve broad-band, near-unit light absorption, which is spatially localised on the nanoparticle layer. 
We show that this enhanced light absorbance leads to $\sim$ 40--fold increases in the photon--to--electron conversion efficiency by the plasmonic nanostructures. 
We developed a model that successfully captures the essential physics of the plasmonic hot--electron charge generation and separation in these structures. 
This model  also allowed us to   establish that efficient hot carrier extraction is limited to spectral regions where the photons possessing energies higher than the Schottky junctions and the localised light absorption of the metal nanoparticles overlap.
}


Metal nanoparticles exhibit absorption cross-sections for incident light that can far exceed their physical dimensions. 
This strong interaction with light originates from the excitation of collective oscillations of surface charges in the nanoparticles, commonly referred to as surface plasmons. 
Plasmons can relax radiatively by re-emitting photons or non-radiatively creating a transient population of non-equilibrium (hot) charge-carriers \cite{Clavero_NP2014a,Brongersma_NN2015a,Linic_NM2011a}, which in a chemically inert environment, transfer their energy to the metal lattice resulting in nanoparticle heating \cite{Hartland_CR2011a}. 
 
However, it is also possible for these hot charge carriers to be deposited into acceptor states of adsorbates \cite{Kale_AC2014a} or to be transferred to an accepting medium creating charge--separated states with sufficient chemical potential energy to drive chemical (redox) reactions.
For instance, in metal-semiconductor Schottky junctions, plasmon relaxation can result in the emission of  hot charge carriers into the conduction band of the semiconductor\cite{Furube_JOTACS2007a}, a charge--separation process that requires photon energies below the band gap energy of the semiconductors and which has been used in photovoltaics \cite{Mubeen_AN2014a}, photodetection \cite{Knight_S2011a} and photocatalysis  \cite{Nishijima_TJOPCL2012a,Tian_JOTACS2005a}.
Plasmonic hot charge--carrier generation  has recently been used for the light-driven 
dissociation of hydrogen\cite{Mukherjee_NL2013a,Mukherjee_JOTACS2014a}, 
oxidation of ethylene\cite{Christopher_NC2011a},
cross--coupling reactions\cite{Xiao_AC2014a}, 
reduction of nitro--aromatics to azo compounds \cite{Ke_GC2013a} 
and other reactions involving more complex organic molecules\cite{Wang_CSR2014a,Scaiano_TJOPCL2013a,Xiao_JMCA2013a}.

The plasmon hot--carrier relaxation pathway is a process that depends on \cite{Moskovits_NN2015a}: 
(i) the efficiency of the charge transfer  that must take place at relevant interfaces, 
(ii) the charge separation following light absorption and 
(iii) the absorption of incident photons.
Therefore, it is expected that the rate of plasmonic hot--charge carrier relaxation can be substantially increased under conditions where the metal nanostructures absorb nearly all incident photons across a broad spectral bandwidth\cite{Fang_NL2015a,Robatjazi_NL2015a}. 
Typically, near unit broadband absorption has been achieved with metal-insulator-metal (MIM)  structures\cite{Liu_AAMI2015a,Aydin_NC2011a,Hedayati_AM2011a,Landy_PRL2008a}.
These  consist of a thin insulator or dielectric layer ($<$ 100 nm) sandwiched between a  metal mirror and a thin top layer of light absorbing metal nanostructures. 
By replacing the insulator layer in an MIM structure with a semiconductor such as TiO$_2$, it should be possible to extract plasmonic hot--carriers using a  Schottky junction that exhibits near unit broadband absorption \cite{Fang_NL2015a}. 
In addition to the expected increases in hot--carrier generation efficiencies,  near--unit broadband absorption is an ideal model system to study the mechanism  of photo--conversion in plasmonic structures.

Here  we present a study on  the extraction of plasmonic hot charge carriers from a monolayer of Au nanoparticles capable of absorbing up to $>$ 90\% of incident visible light in a metal-semiconductor-metal (MSM) configuration.
Examinations of the photoelectrochemical behaviour of such structures reveal significant enhancements in the yields of hot carrier extraction.
Furthermore,  with the aid of a simple phenomenological model, we establish that
plasmon non--radiative relaxation yields uniform energy and momentum distributions of the generated charge carriers, which  limit the spectral bandwidth of  hot carrier extraction.

\begin{figure*}[tbph!]
\centering
\includegraphics[width=0.9\linewidth]{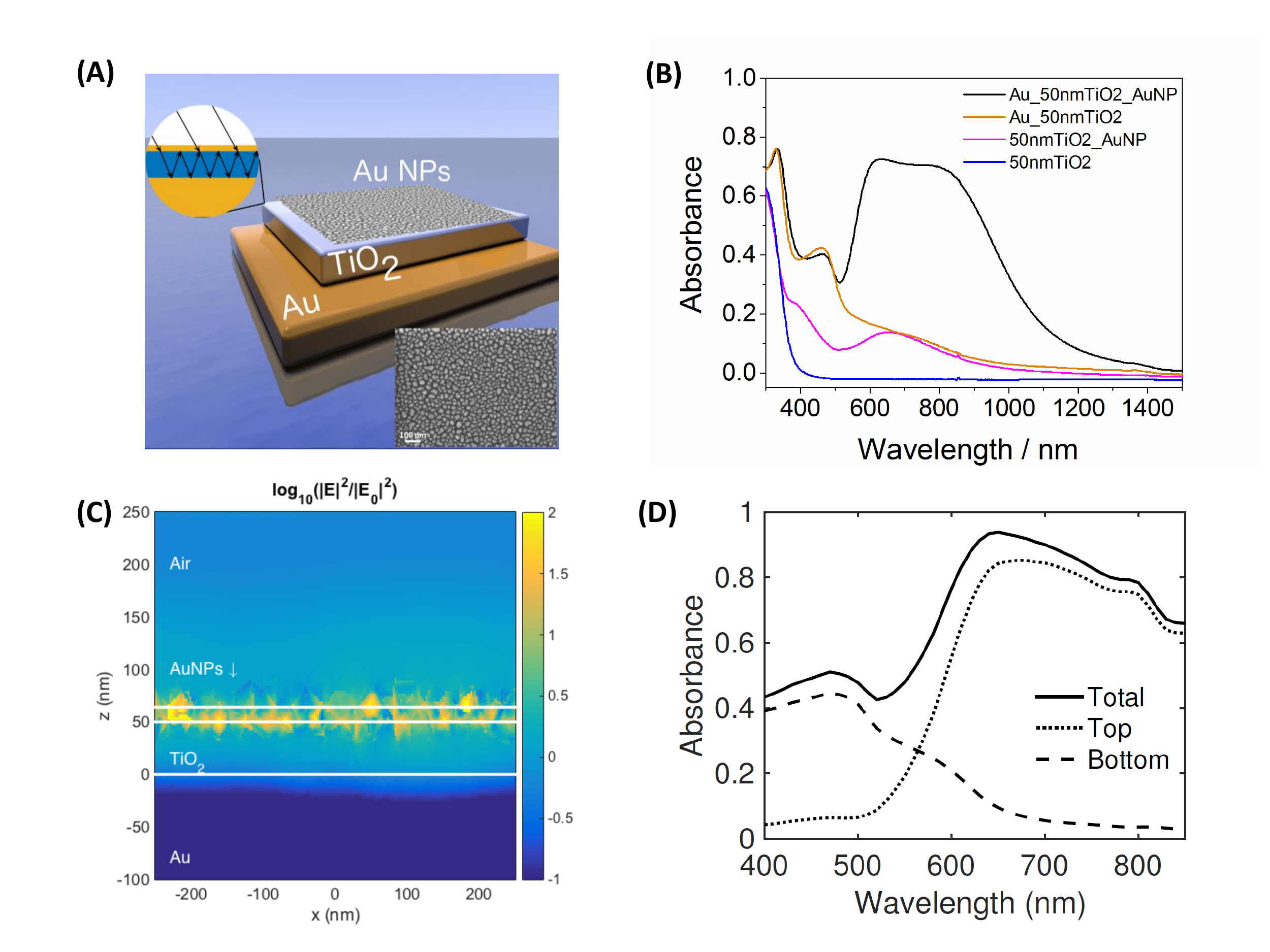}
\caption{\textbf{Architecture and working principle of plasmonic broadband absorber structure}. 
(A) Schematic of the broadband absorber  structure with TiO$_2$ positioned in between the Au  mirror and  Au nanoparticles. 
Inset: multiple internal reflection and interference interpretation of unit absorption \cite{Chen_OE2012a}. 
Also shown  is an SEM of the Au nanoparticle monolayer.
(B) Absorbance spectra of 
Au\textunderscore 50nmTiO2\textunderscore AuNP, 
Au\textunderscore 50nmTiO2 (without Au NPs), 
50nmTiO2\textunderscore AuNP (without mirror) and bare TiO$_2$. 
These spectra were calculated as $A=1-R-T$, where $R$ and $T$ are the measured Reflectance and Transmission spectra (diffuse + specular).
(C) Spatial distribution of the  Electric field  of the metal--semiconductor--nanoparticle structures at 650 nm.
The values are shown on a logarithmic scale of the absolute magnitude square of the electric field relative to the incident field.
(D) Plot of the absorption of incident light {\it vs.} wavelength for the top and bottom metal layers.
}
\label{fig:Fig1}
\end{figure*}

\textbf{Results}

Figure 1a shows the metal--semiconductor--nanoparticle structures, which  consist of an optically thick metal film and \textit{a single layer} of Au nanoparticles, separated by a thin TiO$_2$ layer. 
These samples were made by means of sequential physical vapour deposition of materials (see Methods section),  a large scale method  resulting in a densely--packed monolayer of Au nanoparticles 
[shown in  the scanning electron microscope (SEM) image of Figure 2a: 10 nm thick, average diameter of 20 nm, dimensions that are $\le$ mean-free path for hot electrons in Au\cite{McFarland_N2003a}. Size histogram in  Supplementary figure S1].
This nanoparticle ensemble, when deposited on top of glass--supported TiO$_2$ films,  exhibits a single--band absorption centred around 620 nm (with 0.15 absorbance) characteristic of  localised surface plasmon resonances of Au nanoparticles on materials with a high refractive index \cite{Tian_JOTACS2005a}.
On the contrary, the metal--semiconductor--nanoparticle structure (Au\textunderscore50nmTiO2\textunderscore AuNP) exhibits a distinctively broadband absorption in a spectral region ranging from 600 nm to 1000 nm, reaching absorbance values of up to $\sim$0.75: a 5-fold increase in absorbance, when compared to the control sample without the reflecting layer (50nmTiO2\textunderscore AuNP).
These absorption spectra were calculated as $A=1-R-T$, where $R$ and $T$ correspond to measured reflectance and transmission spectra (both their diffuse and specular components, see Methods section).

In the metal--semiconductor--nanoparticle structures, the thickness of the TiO$_2$ layer was chosen, with the aid of  numerical solutions of Maxwell equations (see supporting information section \ref{sec:S1}) such that the structure exhibited low reflection (R) of light in the visible. 
Given that the thickness of the  supporting metal layer is much greater than the skin depth of the metal, there is no transmission of light in these structures (T = 0) and conservation of energy dictates that the absorbance A is therefore given by A = 1 - R, implying that (near) unit absorption occurs when R (tends to) equals zero. 

Near--unit absorption in these structures originates from optical destructive interference between directly reflected light and the multiple reflections taking place at the other interfaces of the multi--layer stack \cite{Chen_OE2012a} (see diagram of figure 1A).
Other interpretations assign the near--unit absorption of light to a renormalization of the polarizability of the nanoparticles due to their interactions with the mirror \cite{Kwadrin_A2015a} or to optical impedance matching between the structure and free space, a phenomenon that occurs due to a magnetic response  that arises from anti–-parallel currents taking place on both the metal mirror and the nanoparticles \cite{Liu_AAMI2015a,Aydin_NC2011a,Hedayati_AM2011a,Landy_PRL2008a}. 
According to the numerical simulations shown in Figure 1C, at a wavelength of 650 nm there is a strong (and sub--wavelength) localisation of the electric field at the spatial location of the metal nanoparticles, which accounts for the high absorbance reported in Figure 1B.
As a function of the incident photon wavelength, Figure 1D shows that the spatial location of the absorption occurs  at the nanoparticle monolayer for long wavelengths, but it becomes localised on the metal mirror at shorter wavelengths.

\begin{figure}[tbph!]
\centering
\includegraphics[width=0.9\linewidth]{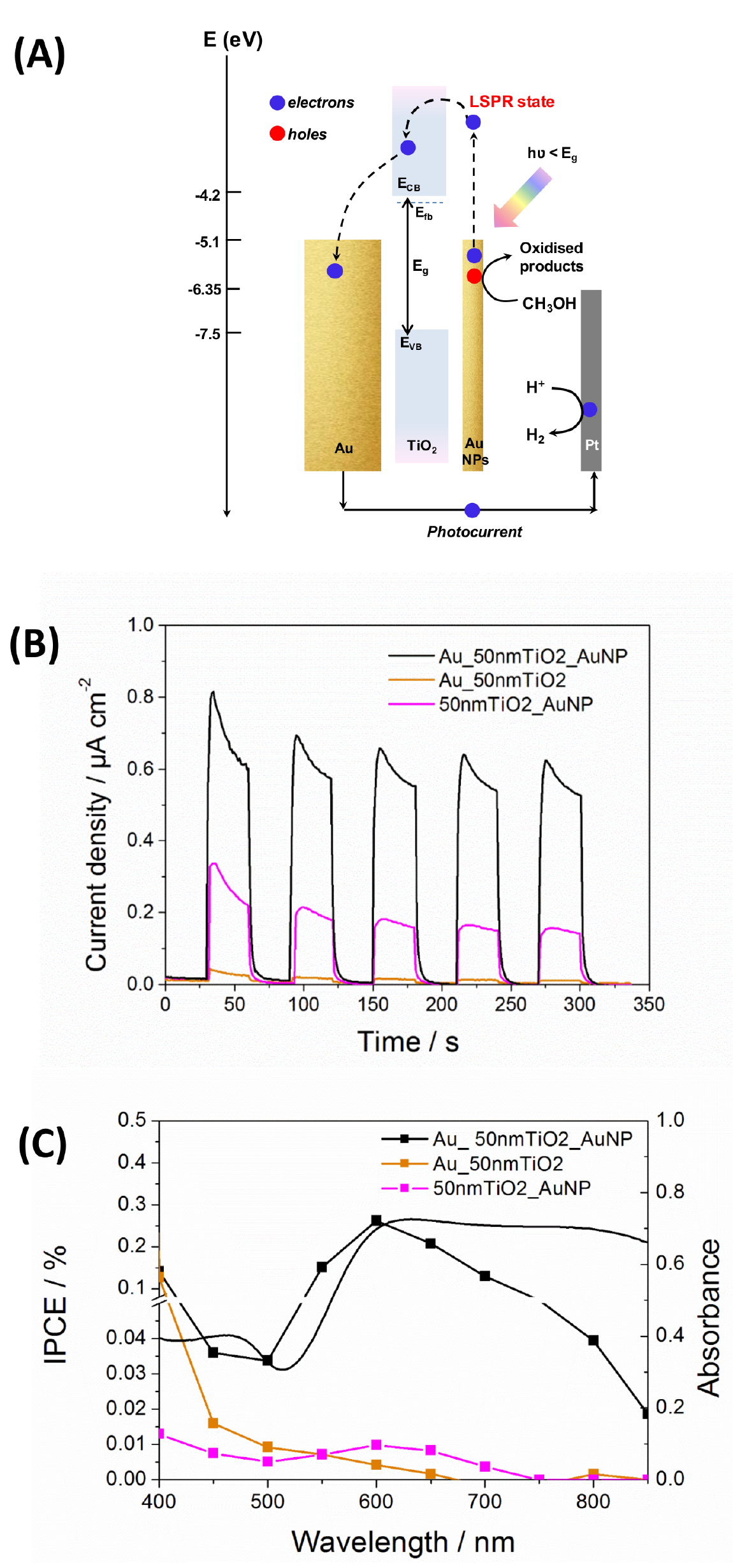}
\caption{\textbf{Photoelectrochemical and absorption performance of the plasmonic broadband absorber structure}.
(A) Illustration of the working principle: Upon visible light illumination and plasmon non--radiative relaxation, hot electrons in the Au nanoparticles can be injected to TiO$_2$ and be transported subsequently to the Pt counter electrode and measured as photocurrents. E$_{CB}$, E$_{VB}$ and E$_g$ refer to conduction band, valence band and band gap energy of TiO$_2$ respectively.
(B) Photocurrent generation under both visible light illumination ($\ge$ 495 nm cut-off filter) and in the dark  in a 2 electrode system vs Pt wire; applied voltage of 0.5 V and 0.5 M Na$_2$SO$_4$ + 20 v/v\% methanol electrolyte solution.
(C) IPCE values  of 
Au\textunderscore 50nmTiO2\textunderscore AuNP, 
Au\textunderscore 50nmTiO2 and 
50nmTiO2\textunderscore AuNP and absorbance data of 
Au\textunderscore 50nmTiO2\textunderscore AuNP.
Also shown is the absorption spectrum of the metal--semiconductor--nanoparticle structure.
}
\label{fig:Fig2}
\end{figure}

\textbf{Photo--currents}.
Subsequent to light absorption, non--radiative relaxation of nanoparticle plasmons can result in the transfer of the incident photon energy to a hot electron, which can obtain sufficient energy and momentum to overcome the Schottky barrier existent at the Au nanoparticle--TiO$_2$ interface \cite{Mubeen_NL2011a}.
To measure this possible photon--to--electron conversion, we employed  the metal--semiconductor--nanoparticle structures as photoanodes in photo--electrochemical cells, as illustrated in the diagram of figure 2A \cite{Thomann_NL2011a,Tian_CC2004a,Garcia-de-Arquer_AN2013a}. 
Electron injection  from the metal nanoparticles into accepting states (such as the conduction band or trap states) of the TiO$_2$ film, leaves a  positive charge (hole) on the Au nanoparticles which can be neutralised by electron donating species in solution.  
Injected electrons in the TiO$_2$ can then be collected by the mirror (which also acts as an electrical contact) and transported to a Pt counter electrode, where reduction reactions take place. 
The overall process results in the generation of measurable photocurrents.

Figure 2B shows the measured photocurrents for three photoanodes:
(i)  metal--semiconductor--nanoparticle  (Au\textunderscore 50nmTiO2\textunderscore AuNP), 
(ii) metal--supported TiO$_2$ thin film (Au\textunderscore 50nmTiO2), 
 and 
(iii) a semiconductor--supported Au nanoparticle monolayer (50nmTiO2\textunderscore AuNP).
These measurements were performed under visible light irradiation (495 nm long pass filter) in a two-electrode configuration with an applied potential of 0.5 V,
and under conditions where the electrolyte consisted of aqueous 0.5 M Na$_2$SO$_4$ with 20 v/v\% methanol (purged with N$_2$, more details can be found in the methods section). 

The long-pass filter  eliminates possible  direct excitation of electron--hole pairs in TiO$_2$ with photon energies above its band gap (see Supplementary figure S2 where we estimate the band-gap of the TiO$_2$). 
Indeed, significant photocurrents at wavelengths higher than 495 nm were only observed when Au nanoparticles were present on the electrodes. 
The photocurrents observed in figure 2B can only be a consequence of the excitation and extraction of hot charge--carriers from the Au nanoparticles into the Pt counter electrode. 
These photocurrents were detected only when the visible light was turned on and the currents returned to the background in the dark. 
This confirms that the measured electrical currents were photo--induced, as opposed to dark (thermal) currents enhanced by the applied bias voltage [see figure S4 where the effect of the applied bias on the measured photocurrents is reported]. 
The measurements were reproducible and stable for all of the samples, indicating the absence of  irreversible electrochemical damage to the photoanodes. 
 The  anodic currents measured with the metal--semiconductor--nanoparticle sample (0.56 $\mu$A cm$^{-2}$) were 3.5 times larger than those obtained with the semiconductor--supported Au nanoparticle monolayer (0.16 $\mu$A cm$^{-2}$) electrode. 
The magnitude of these photocurrents increased linearly with the incident illumination power and illumination leads to a change in the measured photovoltage (see Supplementary figures S3 and S5).

%

\textbf{Incident photon to electron conversion efficiency (IPCE)}.
The IPCE as a function of wavelength was measured for all three samples, and the results are shown in figure 2C. 
The metal--supported TiO$_2$ thin film (Au\textunderscore 50nmTiO2)  did not lead to an appreciable IPCE even though it showed a relatively high absorption at 460 nm. 
On the contrary, the non-vanishing IPCE spectrum measured for the semiconductor--supported Au nanoparticle monolayer (50nmTiO2\textunderscore AuNP) closely follows the lineshape of its corresponding optical absorption spectrum shown in figure 1B, indicating a strong correlation with the photoexcitation of localised surface plasmon resonances. 
For the metal--semiconductor--nanoparticle photo--anode (Au\textunderscore 50nmTiO2\textunderscore AuNP), the measured IPCE has a broadband character with a maximum of 0.27 \% at around 600 nm, an efficiency $\sim$ 20 times larger than the one measured for the semiconductor--supported Au nanoparticle monolayer (50nmTiO2\textunderscore AuNP).

\begin{figure}[tbph!]
\centering
\includegraphics[width=.9\linewidth]{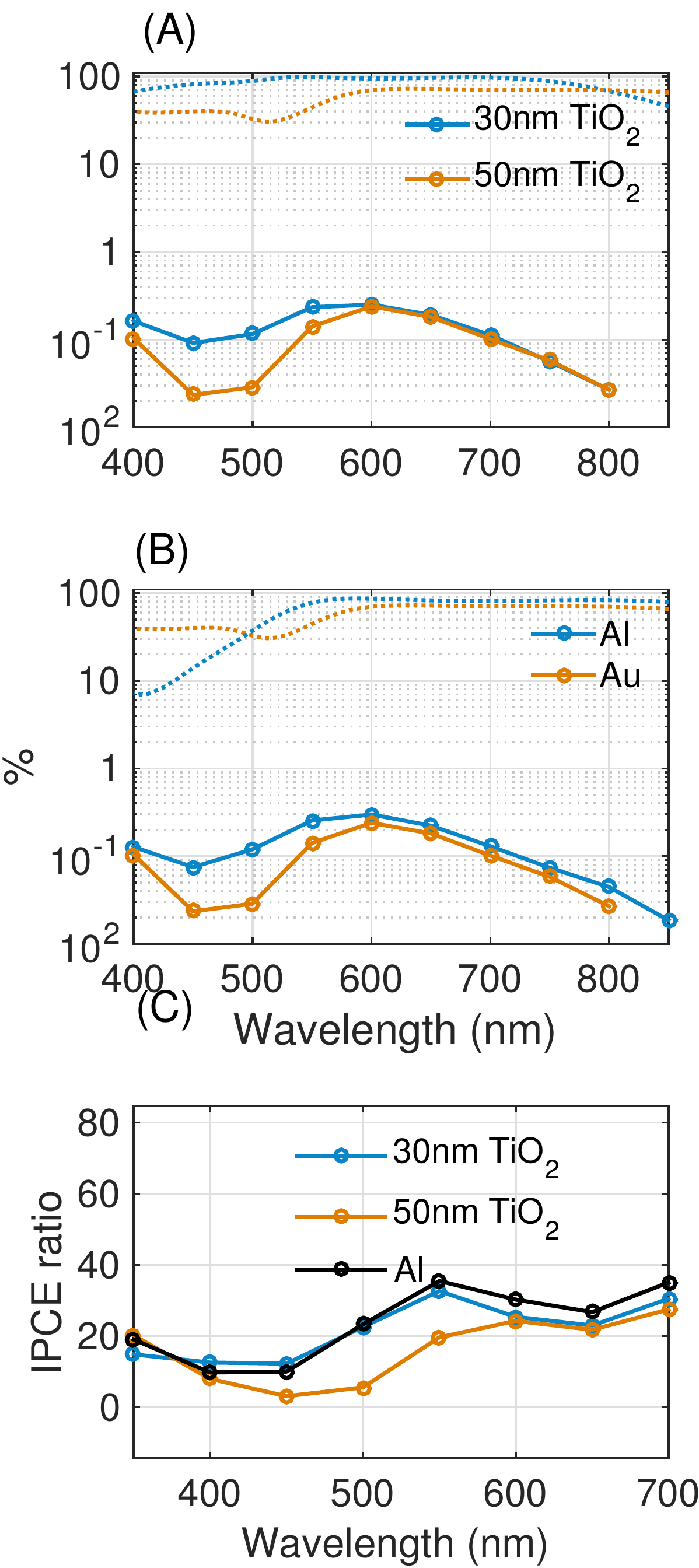}
\caption{\textbf{Effect of structural parameters on measured IPCE.} 
(A)\textit{TiO$_2$ thickness.}  Dashed lines are the measured optical absorption spectra of samples (obtained from diffuse reflectance measurements) with 30 nm and 50 nm of TiO$_2$. Dotted lines show the corresponding IPCE values.  
(B) \textit{Work function of the reflector}. Measured absorption spectra for samples where the metal reflector was made of Au or Al  and their corresponding IPCE spectra (dotted lines). The thickness of TiO$_2$ was kept fixed at 50 nm.
(C) Ratio of measured IPCE relative to the results obtained for the TiO$_2$ supported Au nanoparticle monolayer.
}
\label{fig:Fig3}
\end{figure}
Further increases in the absorbance and (consequently) IPCE values of the metal--semiconductor--nanoparticle structures can be achieved by decreasing the thickness of the TiO$_2$ layer from 50 nm to 30 nm, which as shown in figure 3(A),  leads to an absorbance of up to $\ge$ 95\% (maximum of 99\% at 544 nm). 
The IPCE in turn increases by about a factor of 4$\times$ for wavelengths below 600 nm, but remains almost unchanged at longer wavelengths. 
In figure 3(B), we show that when the  mirror is modified from Au to Al in the metal--semiconductor--nanoparticle stacks (at fixed semiconductor thickness),  
the measured light absorption increases to $\ge$ 80\% (and reaches a maximum of 87\% at 590 nm) with concomitant increases in IPCE values.
Figure 3(C) depicts a summary of the IPCE enhancements  attainable with the  
metal--semiconductor--nanoparticle structures investigated,
which shows increases of almost 40$\times$ with respect to the semiconductor--supported Au nanoparticle monolayer.

 
The measured IPCE spectral lineshapes for the metal--semiconductor--nanoparticle samples track the corresponding optical absorption spectra for wavelengths shorter than $\sim$ 700 nm, but exhibit a characteristic rapid decay at longer wavelengths (in spite of measured strong absorbance) which occurs as a consequence of the mechanism of hot--charge carrier extraction that we proceed to discuss next.

\begin{figure}[tbph!]
\raggedright (A)\\
\centering
\includegraphics[width=.75\linewidth]{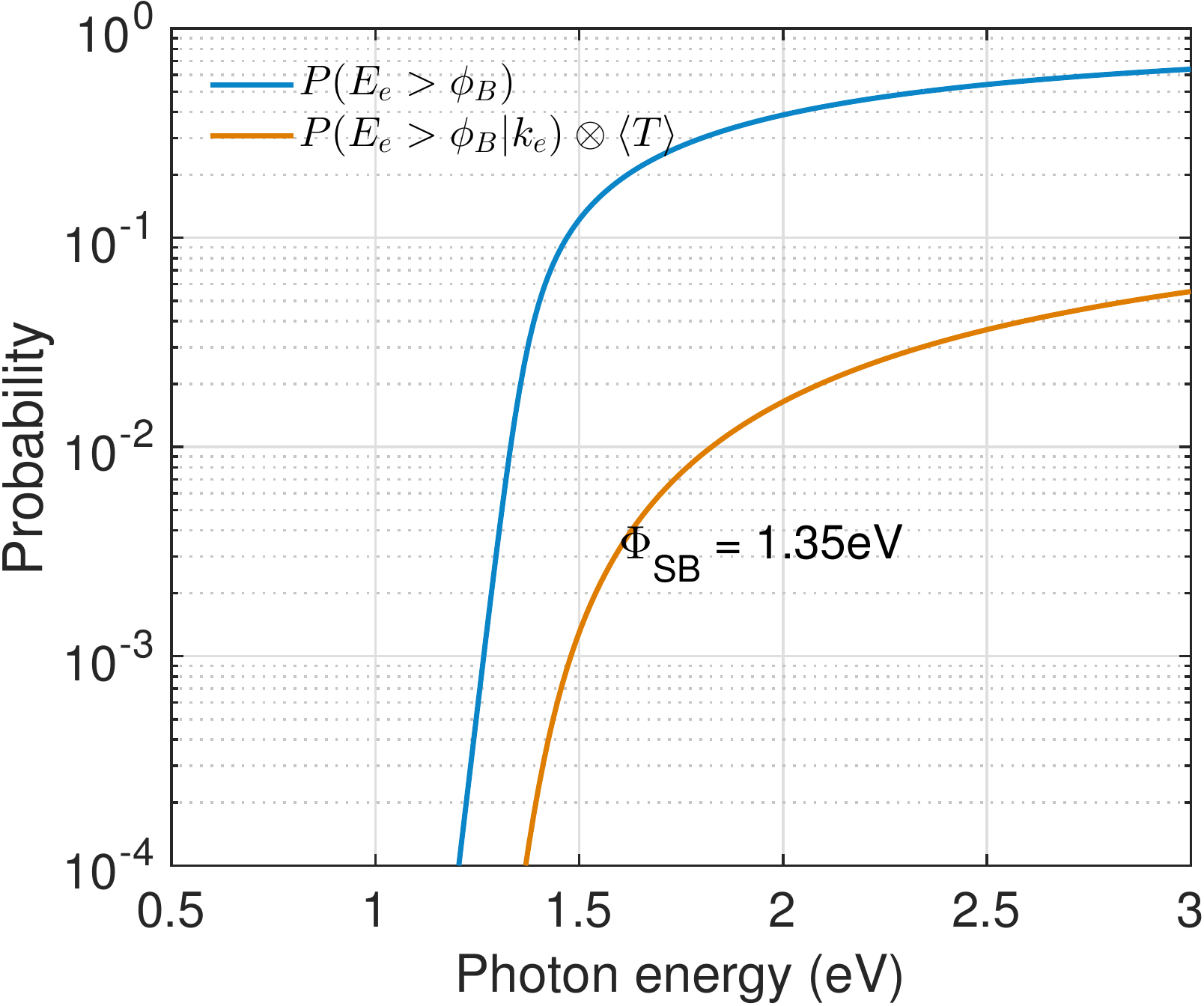}\\
\raggedright (B)\\
\centering
\includegraphics[width=.75\linewidth]{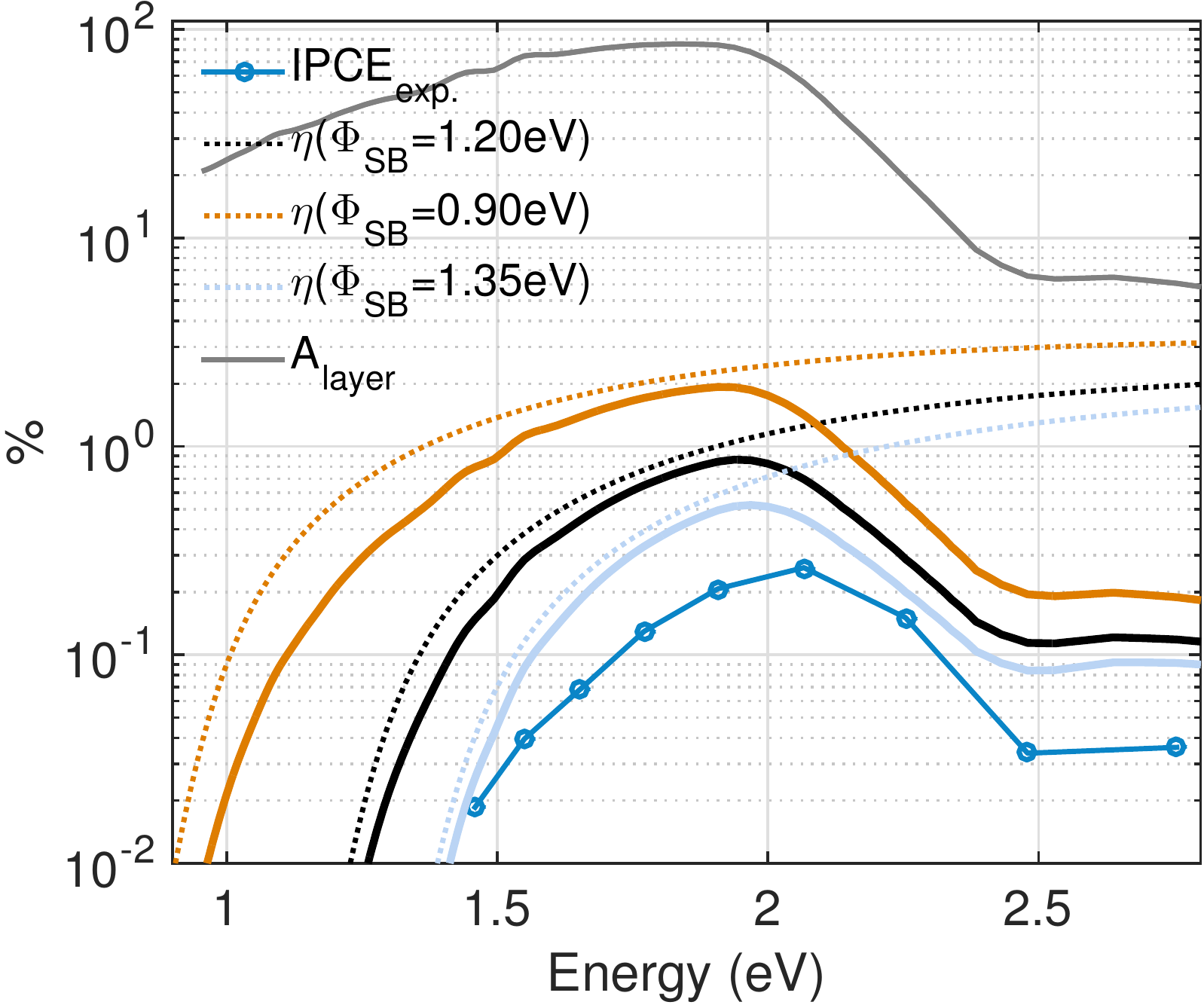}\\
\raggedright (C)\\
\centering
\includegraphics[width=.75\linewidth]{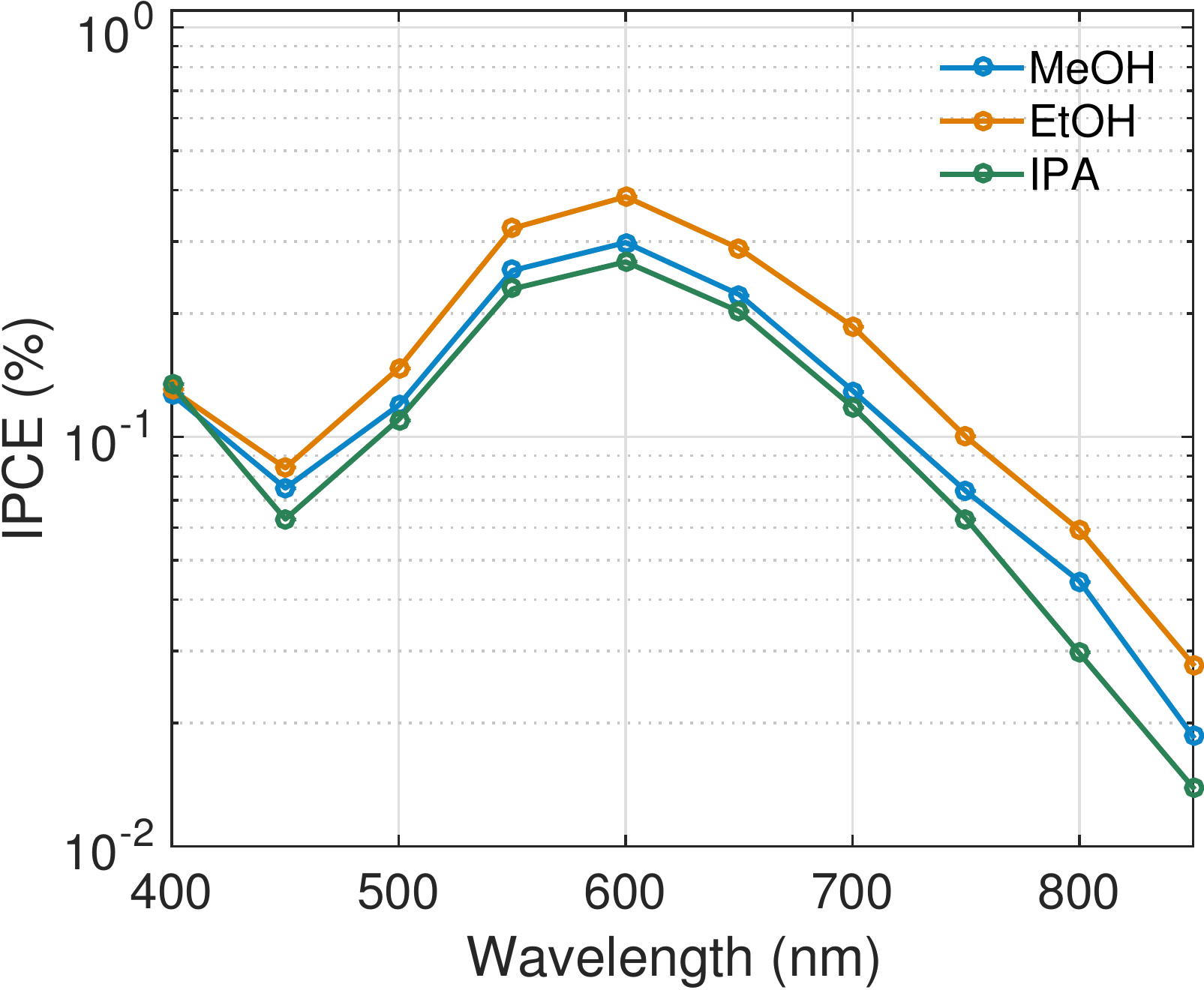}
\caption{ \textbf{Estimate of the efficiency}. 
(A) hot--electron injection probability $\eta_\text{inj} = P(E_e>\Phi_{SB}|k_e)\otimes\langle T\rangle$.  
 $P(E_e>\Phi_{SB}|k_e)$ is 
the joint probability that plasmonic hot--electrons have energies $E_e$ in excess of the Schottky barrier $\Phi_{SB}$ \textit{and} a momentum $k_e$ with a component perpendicular to the metal--semiconductor interface that lies within the escape cone.
$\langle T\rangle$ is the electron transmission coefficient at the nanoparticle--TiO$_2$ interface.
Also shown for comparison, is the probability $P(E_e>\Phi_{SB})$ that the plasmon--derived hot--electrons have energies $E_e$ in excess of $\Phi_{SB}$.
(B) $\eta$: absorbed photon--to--electron conversion efficiency. It shown in this figure for two values of  $\Phi_{SB}$.
The solid curves represent the theoretical estimate of the efficiency given by $A_{layer}\times \eta$, where $A_{layer}$ (dashed grey line) is the calculated absorption at the nanoparticle layer.  
The modelled and measured IPCE (IPCE$_{exp}$) show similar lineshapes exhibiting a pronounced decrease as the wavelength of incident photons approaches $\Phi_{SB}$, consistent with the data shown in (A).
In these calculations  values of 0.96 eV\cite{Zhang_ROPIP2013a}, 1.20 eV\cite{Lee_APL2013a} and 1.35 eV were used for $\Phi_{SB}$. 
(C) Measured IPCE spectra for cases with Methanol (MeOH), Ethanol (EtOH) and Isopropanol (IPA) as the sacrificial electron donors in solution.}
\label{fig:Fig4}
\end{figure}

\textbf{Discussion}

The IPCE at a specific wavelength $\lambda$ is proportional to the photon absorption efficiency  of  the metal nanoparticles $A_{layer}$ and the probability $\eta$ that the absorbed photon results in a collected electron in the photoelectrochemical circuit\cite{McFarland_N2003a}:
\begin{equation}
\text{IPCE}(\lambda) = A_{layer}(\lambda)\times\eta(\lambda).
\end{equation}
The former can be determined by solving Maxwell's equations for the structures, which also provide information on, for instance,  the spectral and spatial location of the light-absorption events (e.g. figure 1D). 
 $A_{layer}$  is optimal for   structures that exhibit unit absorption of light at the position of the nanoparticle monolayer  (i.e. $A_{layer}=1$). 

In order to get a quantitative estimate for $\eta$, as a first--order approximation, this parameter  can be expressed as (more details in Supplementary section S3):
\begin{equation}\label{eq:eta}
\eta = \eta_\text{inj}\times \eta_\text{trpt}\times\eta_\text{injm}\times\eta_{ed},
\end{equation}
which conceptually 
results from the following sequence of  events (see the diagram shown in figure 2): 
 	(i) the  injection of a plasmon--derived hot electron across the nanoparticle--semiconductor Schottky barrier, with probability  $\eta_{inj}$ ,
 	(ii) elastic transport of the injected electrons across the semiconductor layer with an associated probability $\eta_{trpt}$, 
 	(iii) the injection of the charge carrier into the metal reflector, occurring with a probability $\eta_{injm}$. 
 	An additional assumption is made in that once injected into the reflector, the charge carrier will travel through an external circuit to the counter electrode  where finally 
 	(iv) the total current is measured when an electron--donating species in solution injects an electron into the positively--charged metal nanoparticles with a probability $\eta_{ed}$ \cite{Moskovits_NN2015a}.

The hot--electron injection efficiency $\eta_{inj}$ is determined by the  energy--momentum distribution of the hot--electron population that results from Landau damping of plasmons.
Little is known at present about these  energy--momentum distributions.
To get an estimate of $\eta_{inj}$, we assume that Landau damping produces an isotropic momentum distribution \cite{Scales_QEIJO2010a} and approximate the energy distribution of hot--electrons as a product of the initial and final (parabolic) density of electron states in the metal.
We furthermore consider the initial states to have energies  ranging from below (and up to) the Fermi level $E_F$, whereas the final states are thought to have energies   from the Fermi level and up to  $E_F$ + the incident photon energy \cite{White_APL2012a}.
$\eta_{inj}$ is then approximated as  $P(E_e>\Phi_{SB}|k_e)\otimes\langle T\rangle$ [see equation \eqref{eq:S_eta_inj}], where: 
$\langle T\rangle$ is the transmission coefficient for electrons across the metal nanoparticle--semiconductor interface (calculated taking into account conservation of momentum, see section \ref{sec:S_T})
and,
$P(E_e>\Phi_{SB}|k_e)$ is the  fraction of the   hot--electron population for which the electrons have  a momentum $k_e$ that lies within the escape cone of the metal--semiconductor interface and 
with energies $E_e$ above the Schottky barrier [see equation \eqref{eq:S_momentum}].
As shown in figure 3A,  $\eta_{inj}$ increases  slowly with incident photon energy
[details of the calculations leading to figure \ref{fig:Fig4}(A) are shown in the supplementary section S3, note that we do not consider possible quantum mechanical tunnelling of electrons under the barrier].


The charge--carrier transport efficiency $\eta_\text{trpt}$ is the probability that, after injection, the charge carrier is transported    through the TiO$_2$ layer, without experiencing  scattering and/or trapping at defect states, the probability of which we assume to be determined by the ratio of the thickness of the metal--oxide layer  and the  mean--free path for electrons\cite{Seah_SAIA1979a} (see section \ref{sec:eta_trpt}).
$\eta_{injm}$, is approximated by the transmission coefficient for  electrons to traverse the semiconductor--metal (reflector) interface, thus accounting for possible reflections 
due to the momentum and energy mismatch  
(more details in section \ref{sec:S_eta_injm}). 
For simplicity, we do not account for possible electron flow from the metal reflector into the semiconductor, which may originate from the excitation and decay of surface plasmon polaritons at the metal/semiconductor interface.
This charge flow will have an opposite direction and thus decrease the measured photocurrents and IPCE \cite{Chalabi_NL2014a, Wang_NL2011a}.

The calculated values of $\eta$ (assuming $\eta_{ed}=1$) and 
$A_{layer}\times\eta$ produce the dotted and continuous lines in Figure \ref{fig:Fig4}B respectively.
$\eta$  is strongly dependent on the magnitude of $\Phi_{SB}$, the thickness of the oxide layer and the height of the energy barrier preventing electron injection at the semiconductor--mirror interface.
The model satisfactorily reproduces the slow increase of the measured IPCE with incident photon energy, in particular, for a value of $\Phi_{SB}$ = 1.20 eV reported by the experiments of Lee {\it et al} \cite{Lee_NL2011a} 
In the model, this curvature originates from a combination of two factors:
(i) the almost rectangular  shape of the energy distribution of hot carriers, which  limits the number of these carriers that meets the energy requirements for injection [i.e. $E>\Phi_{SB}$, see curve $P(E_e>\Phi_{SB})$ of fig. 4A] and,
(ii) the narrow escape cone for carriers from the metal nanoparticles and into the semiconductor, which  severely limits the total fraction of ``useful'' charge carriers by almost 75\% [see figure S9(B)].
Factor (i) is a consequence of the assumed broad energy distribution of initial states, whereas (ii) originates from the assumed uniform distribution of electron momenta resulting from Landau damping.
At higher energies, the calculated drop in $A_{layer}\times\eta$ is in qualitative agreement with the experimental results, which is a consequence of the fact that at these wavelengths, light is predominantly absorbed by the mirror (see figure 1D).
A simple conclusion that can be drawn from these observations is that, in spite of the spectrally broad and strong absorption of light achievable with the metal--semiconductor--nanoparticle structure, the generation of charge--carriers from plasmon relaxation in this architecture is limited to those spectral regions where  the incident  photon energies are  in excess of  the metal--semiconductor Schottky barrier and below the onset for strong absorption by the metal reflecting layer.

\begin{figure}[tbph!]
\centering
\includegraphics[width=0.9\linewidth]{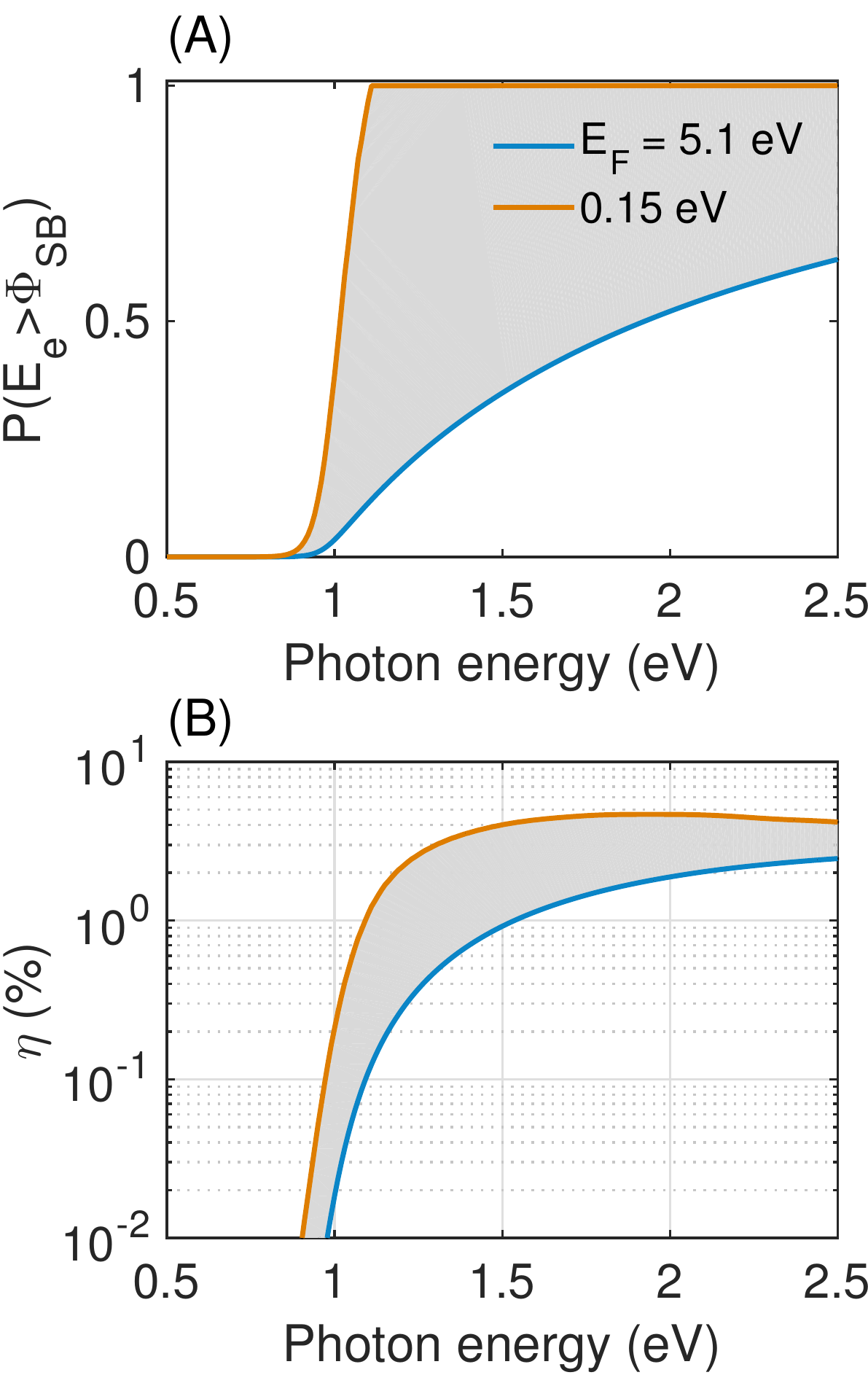}
\caption{\textbf{Narrow distribution of states near the Fermi level of the metal}. (A) Distribution of electron energies and  (B) calculated internal quantum efficiency.
 The shaded areas highlight the attainable differences. $\Phi_{SB}$ = 0.98 eV.}
\label{fig:5}
\end{figure}
According to the model, the ICPE is expected to increase if the  plasmon relaxation were to result in narrow electron energy distributions.
This is a situation that can be modelled by considering that the bottom of the conduction band is close ($\sim$ 0.15 eV) to the Fermi level of the metal \cite{Chan_QEIJO1980a, White_APL2012a}, a hypothetical (and somewhat unrealistic) situation depicted in 
figure \ref{fig:5}.
The orange curve in fig. \ref{fig:5}(A) shows that for this hypothetical case, the fraction of the population of hot--electrons with $E_e>\Phi_{SB}$ is almost a step function of the incident photon energy and becomes energy--independent (and unit) for energies slightly above $\Phi_{SB}$.
This step--like $P(E_e>\Phi_{SB})$ drastically changes the shape of $\eta$ [fig. \ref{fig:5}(B)] and leads to an overall increase in the expected internal efficiency

The measured IPCEs are  well below the predictions of our model, which were made under the assumption of unit $\eta_{ed}$.
$\eta_{ed}$ in the photo--electrochemical configuration that we adopted for extracting hot--charge carriers is
expected to be limited by the kinetics of charge and mass transfer occurring at the solid--liquid interfaces.
In figure 4C we show the measured IPCE changes for a sample where the (sacrificial) electron donating species were changed from Methanol  (MeOH) to Ethanol (EtOH) and to Isopropanol (IPA).
An uniform increase in the IPCE is clearly observable for the case of EtOH, which is attributable solely to changes in the kinetic and thermodynamic processes taking place at the photoanode/liquid junction.
$\eta_{ed}$ is a parameter which is not included in our theoretical description of plasmonic hot--charge carrier extraction, but which seems to play a key role in determining the IPCE in our photo--electrochemical cell, an observation also conjectured in (solid state) plasmonic solar cells\cite{Reineck_AM2012a}, where it was found that hole accumulation at the metal nanoparticles limits device performance.

In summary, we have demonstrated the extraction of plasmonic--derived charge carriers from a multi--layer stack comprising a monolayer of Au nanoparticles.
These multilayer structures exhibit broadband and intense 
absorption of light, which leads to a significant increase in the incident photon--to--electron conversion efficiencies.  
We have developed a simple model to describe the hot--charge carrier generation and transport in this system, which satisfactorily describes the measured IPCE spectra.
According to this  model, the broad distribution of hot--charge carrier energies and their uniform distribution of momenta, account for the weak increase in measured IPCE with  incident  photon energies.
A more efficient strategy for generating and extracting plasmonic hot charge carriers is one that combines strong near--unit absorption of light with a mechanism, such as the plasmon-induced metal-to-semiconductor interfacial charge transfer transition, where a quantum yield for electron injection of $>$24\% has been found to be  independent of the incident photon energy \cite{Wu_S2015a}. 
Our results demonstrate that significant enhancements in
the efficiencies of opto--electronic and photoelectrochemical devices that operate with plasmonic--derived hot--charge carriers can be achieved by tailoring the optical properties of the plasmonic nanostructures.

\textbf{Methods}

{\footnotesize  

\textbf{Material fabrication and Characterization}
The Al and Au  mirrors were deposited onto glass substrates with an in--house constructed electron beam evaporator equipped with transmittance monitoring. 
A thin layer of Cr was deposited prior to the Au deposition to ensure good adhesion to the glass substrate. 
For both Al and Au mirrors, deposition was conducted by monitoring the optical transmittance  to ensure the films had no transmittance of light. 
Subsequently, the TiO$_2$ layers (30 nm and 50 nm) were deposited onto the metal supporting mirrors through an ion--assisted electron beam evaporation process at 200$^\circ$C. 
The bombardment of oxygen during the deposition process ensured the deposition of stoichiometric TiO$_2$ onto the mirrors.  
Lastly, the Au nanoparticles were deposited with a similar electron beam evaporation process with two types of in--situ optical monitoring. 
A broadband transmittance was utilized to monitor the absorption of the Au nanoparticles, while an ellipsometry measurement at 633 nm was employed to achieve non--overlapping particles and detect the deposition point at which  the particles coagulate to form continuous films. 
Scanning electron microscope (SEM) images of the Au NPs were obtained with a Zeiss Merlin Field Emission Scanning Electron Microscope. 
The diffuse and specular reflectance ($R$) and  transmittance ($T$) spectra were measured using a UV-VIS spectrophotometer (Perkin-Elmer Lambda 1050) with an integrating sphere and small spot kit. 
With these two measurements, the absorbance was calculated as $A = 1 - R - T$. 

\textbf{Photoelectrochemical measurements}
The designed structure and platinum wire was employed as the working electrode and counter electrode respectively in a 2 electrode system. 
The exposed surface area of the working electrode was 1 cm in diameter. 
The electrolyte solution used was 0.5 M Na$_2$SO$_4$ (anhydrous, Sigma-Aldrich, $\ge$99\%) with 20 v/v \% amount of methanol (Sigma-Aldrich, $\ge$99.9\%) or ethanol (Sigma-Aldrich, $\ge$99.9\%) as the sacrificial reagent.
 The 0.5 M Na$_2$SO$_4$ was employed to reduce the resistance in the electrolyte solution. 
 Prior to the photocurrent measurements, the solutions were purged with N$_2$ gas to remove  electron scavenging O$_2$. 
 The working potential was set at +0.5 V versus the Pt wire and the devices were illuminated with light from a 300 W Xenon lamp (Newport Model no. 669092) using a $\ge$ 495 nm cut-off filter (Thorlabs FGL-495). 
 The applied voltage and photocurrent $I$ were recorded with a potentiostat (AutoLab PGSTAT204). 
 To measure the IPCE values, light from the Xenon lamp was coupled to a monochromator with a bandwidth of 5 nm at full width at half maximum (FWHM). 
 The light intensity $P$ of the monochromatic light at each wavelength was measured using a Thorlabs optical power and energy meter (Model. PM100D). 
 The incident photon to electron efficiency (IPCE) was calculated by the following formula: 
\begin{equation}
\text{IPCE} = 100\times\frac{I(A/cm^2)}{P(W/cm^2)}\times\frac{1240}{\lambda(nm)}.
\nonumber
\end{equation}

} 
\cleardoublepage


\vspace{0.5cm}

\textbf{Acknowledgments} This work was performed in part at the Melbourne Centre for Nanofabrication (MCN) in the Victorian Node of the Australian National Fabrication Facility (ANFF).
C. N. was supported by an OCE Fellowship from CSIRO.
D.E.G. acknowledges the ARC for support through a Future Fellowship (FT140100514) 
D.E.G. and T.J.D. acknowledge the ANFF for the MCN Technology Fellowships.

\textbf{Author contributions}
  C.N. and S.D. prepared the samples. C.N. performed all the experiments. 
  J.C. and A.R. performed the numerical simulations.  
  All the authors contributed to writing the manuscript.

\textbf{Additional information:}
\textbf{Supplementary Information} accompanies this paper 

\textbf{Competing financial interests:} The authors declare no competing financial interests.

\cleardoublepage
\clearpage
\appendix

\renewcommand{\thesection}{S\arabic{section}}
\renewcommand{\thesubsection}{.\arabic{subsection}}
\renewcommand{\thefigure}{S\arabic{figure}}
\renewcommand{\theequation}{S\arabic{equation}}
\renewcommand{\thetable}{S\arabic{table}}
\setcounter{equation}{0}
\setcounter{figure}{0}

\section{Additional supporting figures}
\label{sec:S1}
\begin{figure}[tbph!]
\centering
\includegraphics[width=0.7\linewidth]{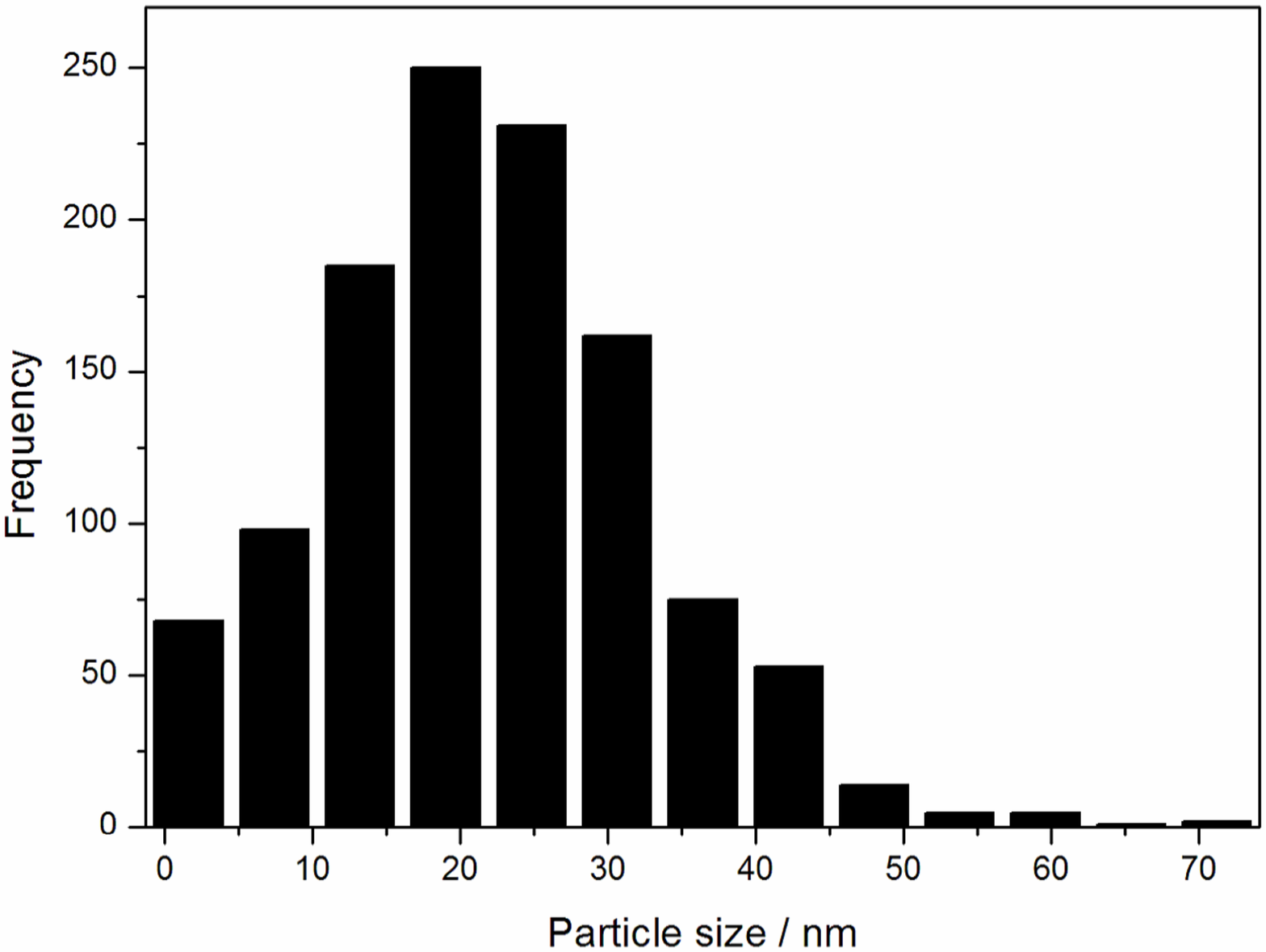}
\caption{Measured size distribution of the  Au nanoparticles.}
\label{fig:FigS1}
\end{figure}

\begin{figure}[tbph!]
\centering
\includegraphics[width=0.7\linewidth]{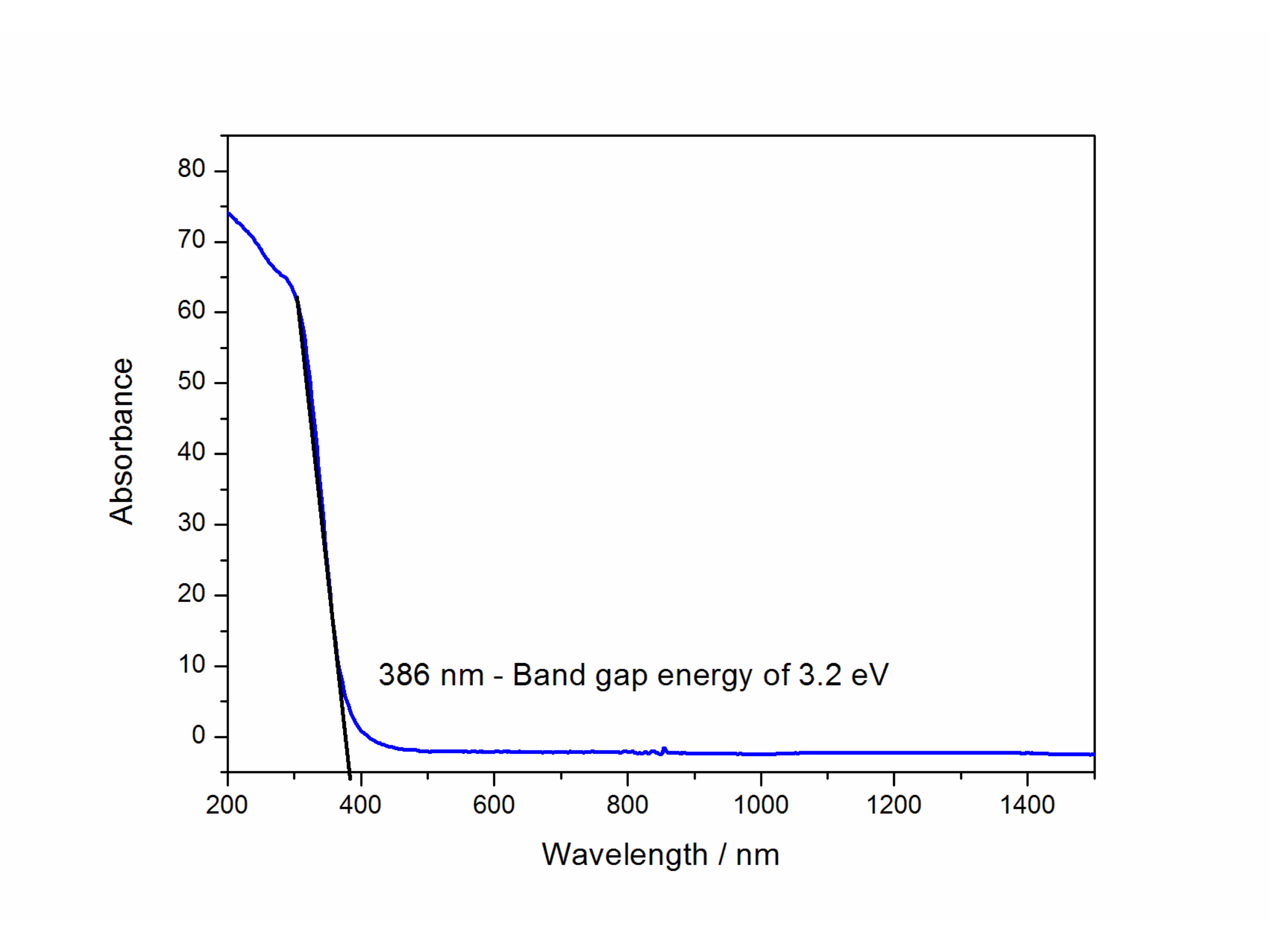}
\caption{Absorption spectrum of the bare TiO$_2$ film. Analysis of these data yields a bandgap of 3.2 eV for the material as indicated.}
\label{fig:FigS2}
\end{figure}

\begin{figure}[tbph!]
\centering
\includegraphics[width=0.7\linewidth]{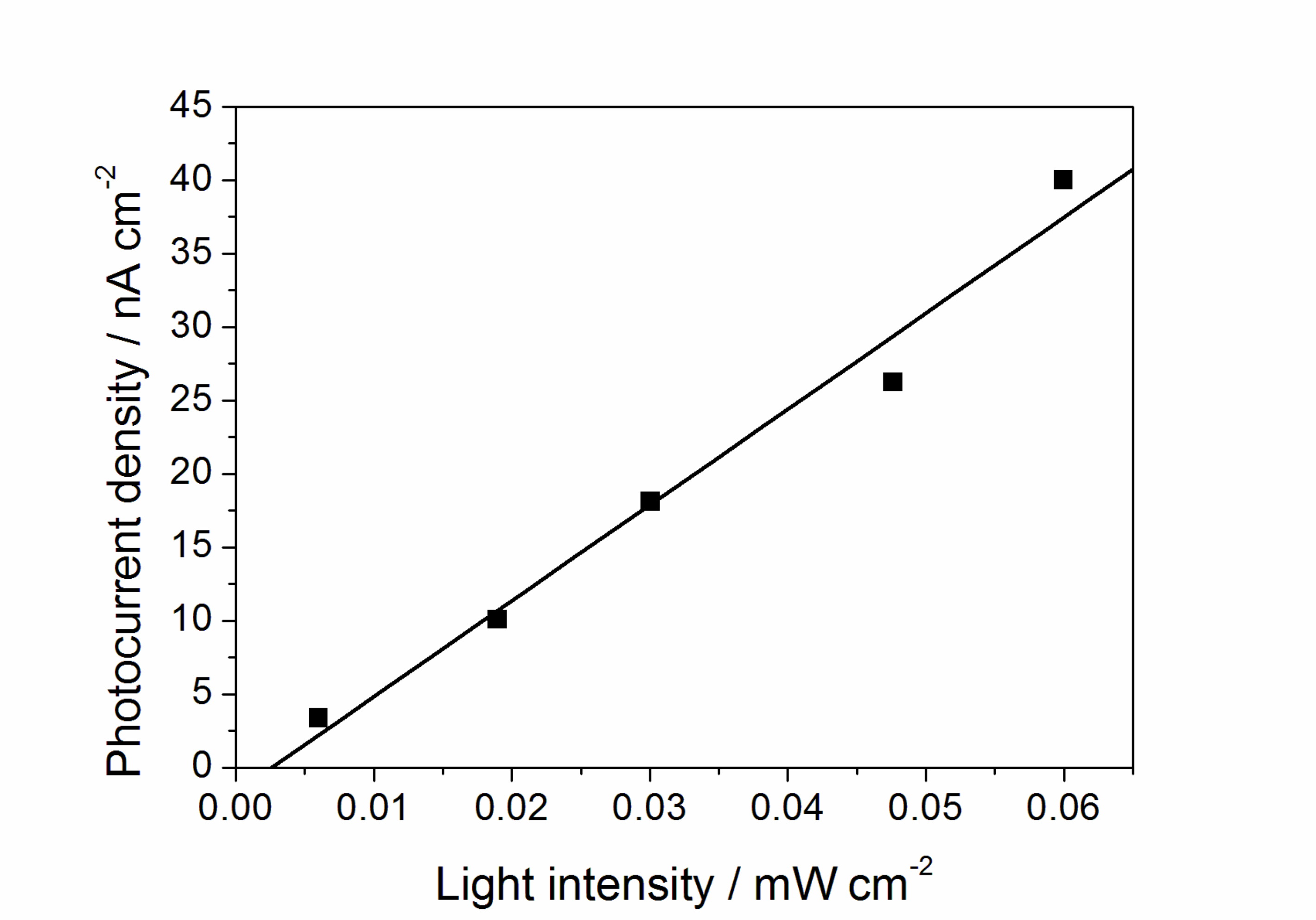}
\caption{IPCE vs light intensity. Sample: Al/50nmTiO2/AuNP
Electrolyte: 0.5M Na$_2$SO$_4$ with 20v/v\% methanol
Wavelength: 600 nm. Bias Voltage: 0.5 V}
\label{fig:FigS3}
\end{figure}

\begin{figure}[tbph!]
\centering
\includegraphics[width=0.7\linewidth]{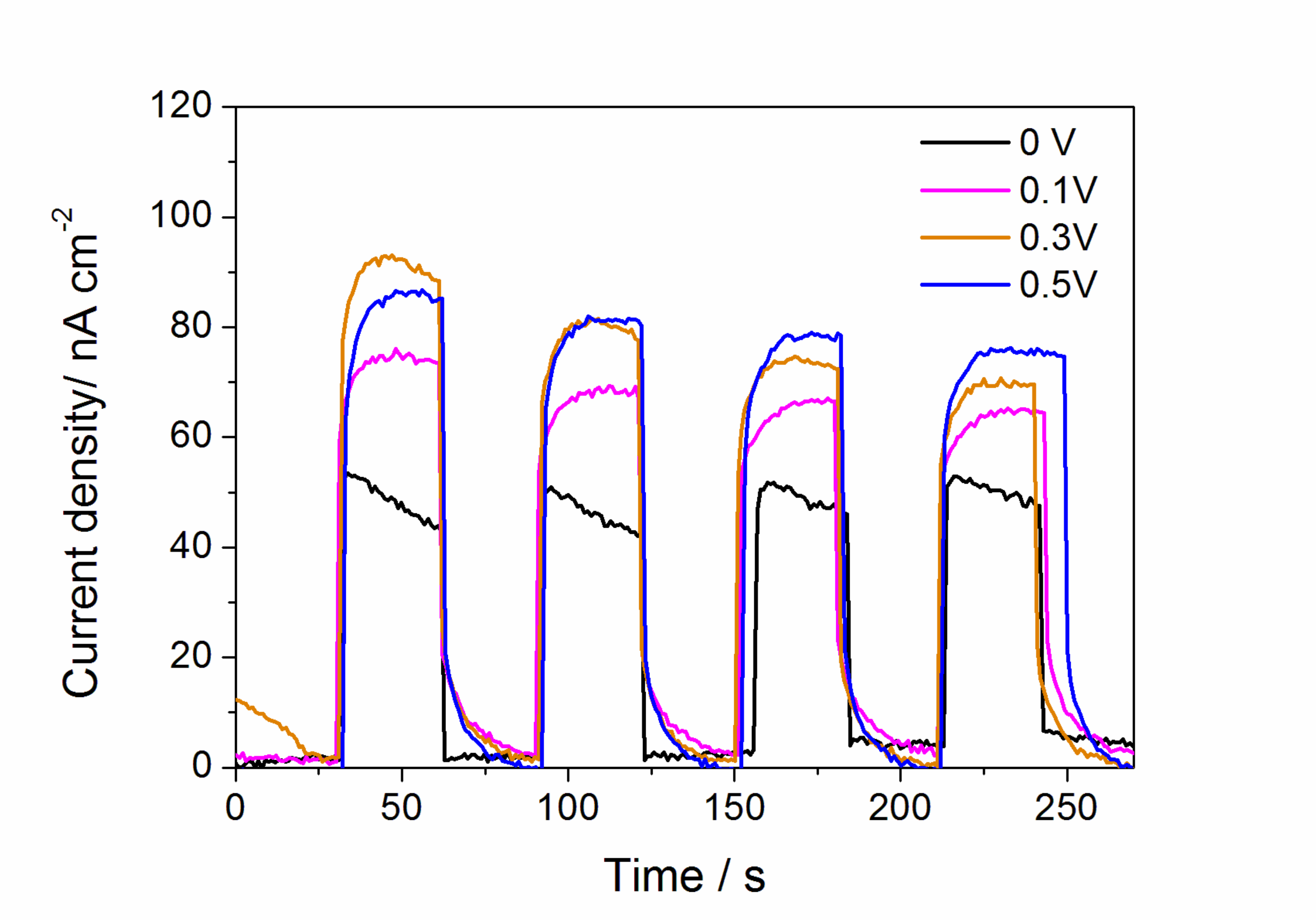}
\caption{IPCE vs light intensity. Sample: Al/50nmTiO2/AuNP
Electrolyte: 0.5M Na$_2$SO$_4$ with 20v/v\% methanol
Wavelength: 600 nm. Bias Voltage: 0.5 V}
\label{fig:FigS4}
\end{figure}

\begin{figure}[tbph!]
\centering
\includegraphics[width=0.7\linewidth]{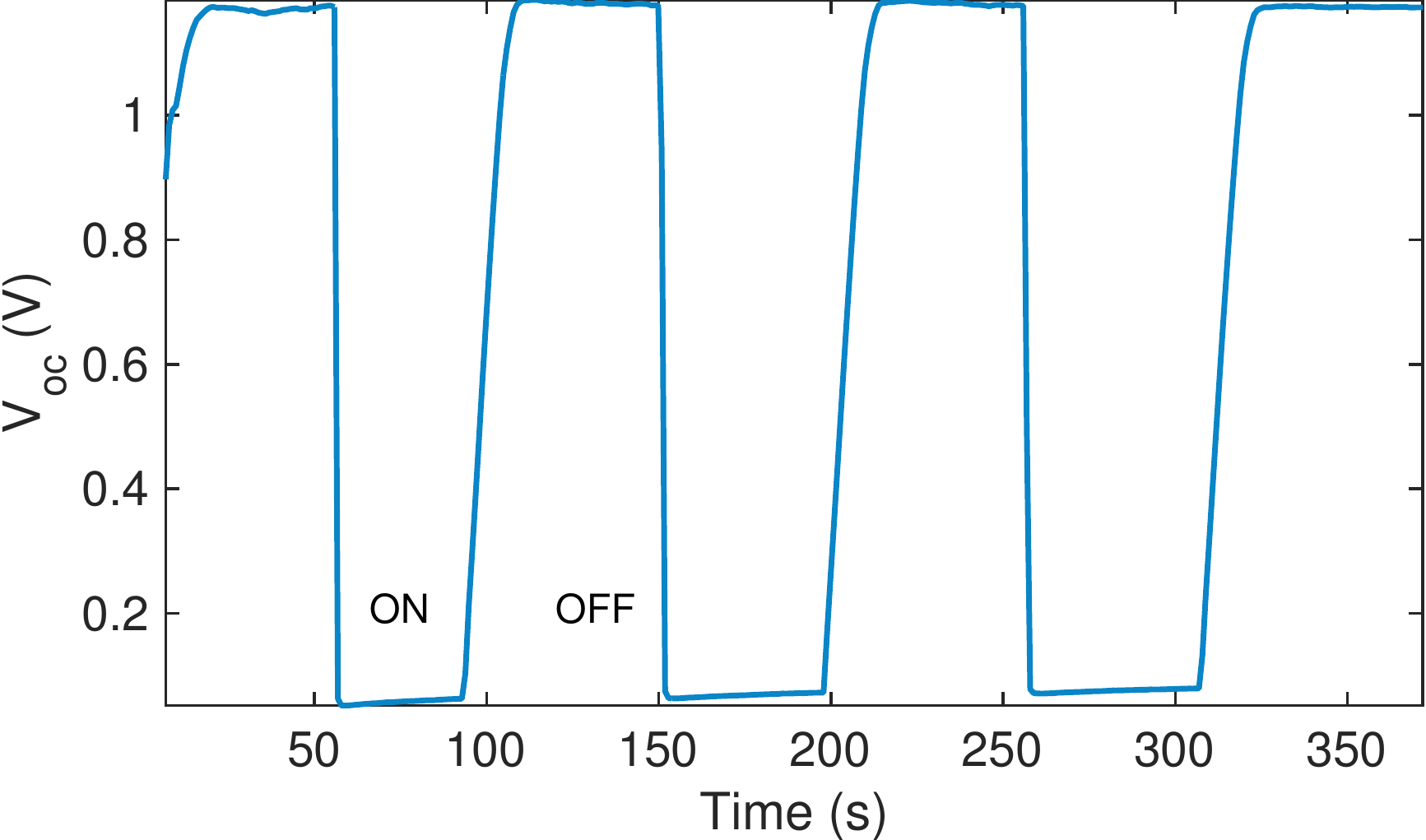}
\caption{Time evolution of the measured open--circuit voltage vs a Ag/AgCl reference electrode. The photo--anode consisted of Au/TiO$_2$[50 nm]/Au. Electrolyte: 0.5M Na$_2$SO$_4$ with 20v/v\% methanol. Light source 300 W Xenon lamp (Newport Model no. 669092) using a $\ge$ 495 nm cut-off filter (Thorlabs FGL-495). }
\label{fig:Voc_vs_t}
\end{figure}

\cleardoublepage
\section{Finite element method (FEM) calculations }
\label{sec:S2}
\begin{figure}[tbph!]
\centering
\includegraphics[width=0.9\linewidth]{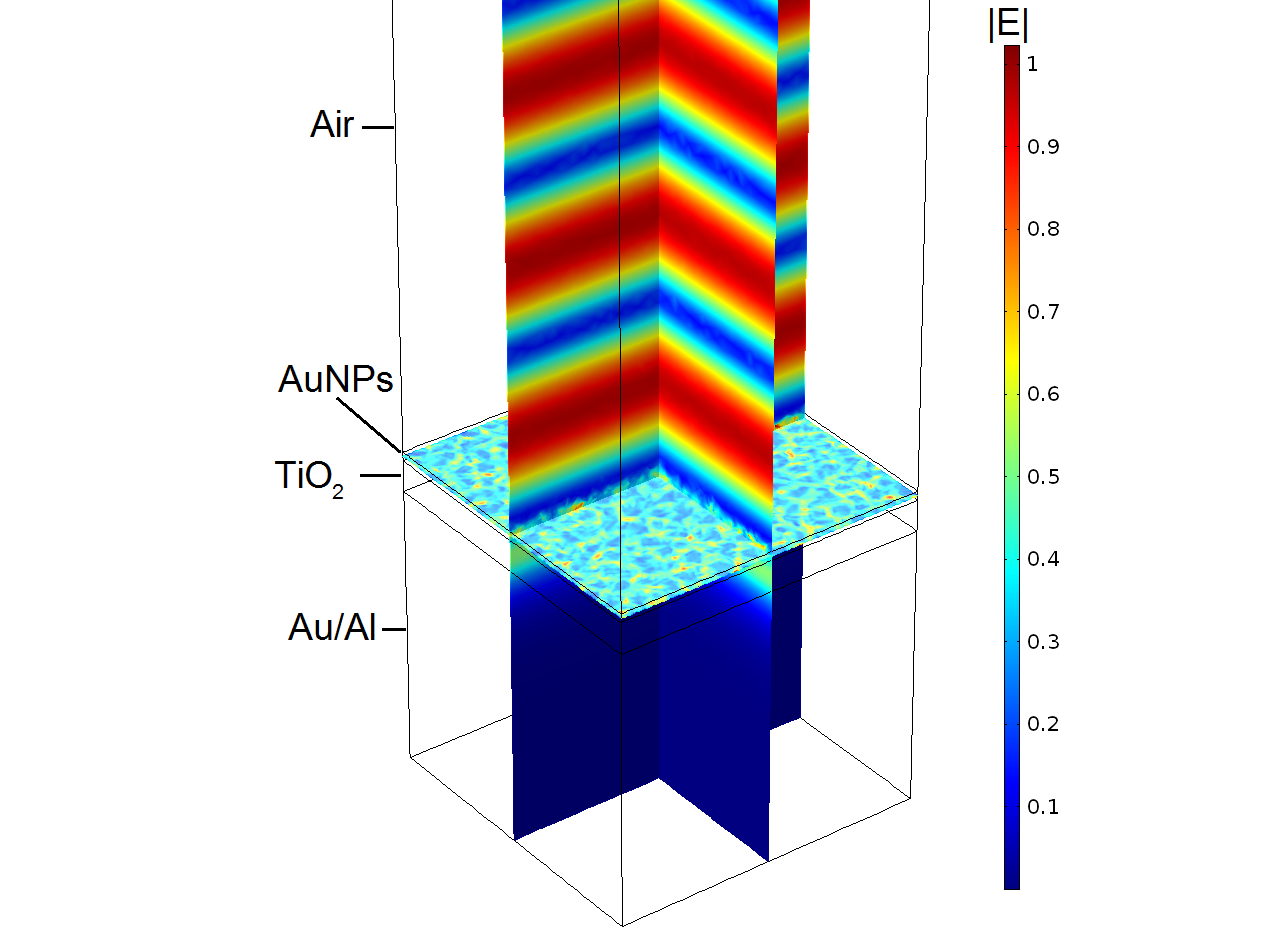}
\caption{The FEM model geometry used to calculate the total absorption of the device and the location of the absorption. Here the normalized magnitude of the electric field is shown. }
\label{fig:geometry}
\end{figure}

The FEM calculations were carried out using COMSOL Multiphysics 5.0, using the model geometry shown in  
figure \ref{fig:geometry} (which also shows the the electric field). 
The model consists of a  500 nm x 500 nm x 2000 nm unit cell, with periodic boundary conditions on the sides and scattering boundary conditions for the top and bottom boundaries. 
The nanoparticle (AuNP) layer was 14 nm thick, the TiO$_2$ layer was allowed to have variable thickness and the reflecting mirror had a fixed thickness of 150 nm.
The refractive index of Au taken from literature \cite{Palik_1985a}. 
The AuNP layer thickness was determined by calculating the surface coverage of nanoparticles (47\%) from the SEM image shown in the main text and applying volume conservation.
A 7 nm solid gold film has the same volume as a 14 nm film with 47\% surface coverage, so we take 14 nm as the average height of the nanoparticles. 
The optical constants of the TiO$_2$ film  were taken from ellipsometry data.
Illumination was modelled with a plane wave launched from the top boundary of the simulation geometry. 

\begin{figure}[tbph!]
\centering
\includegraphics[width=0.9\linewidth]{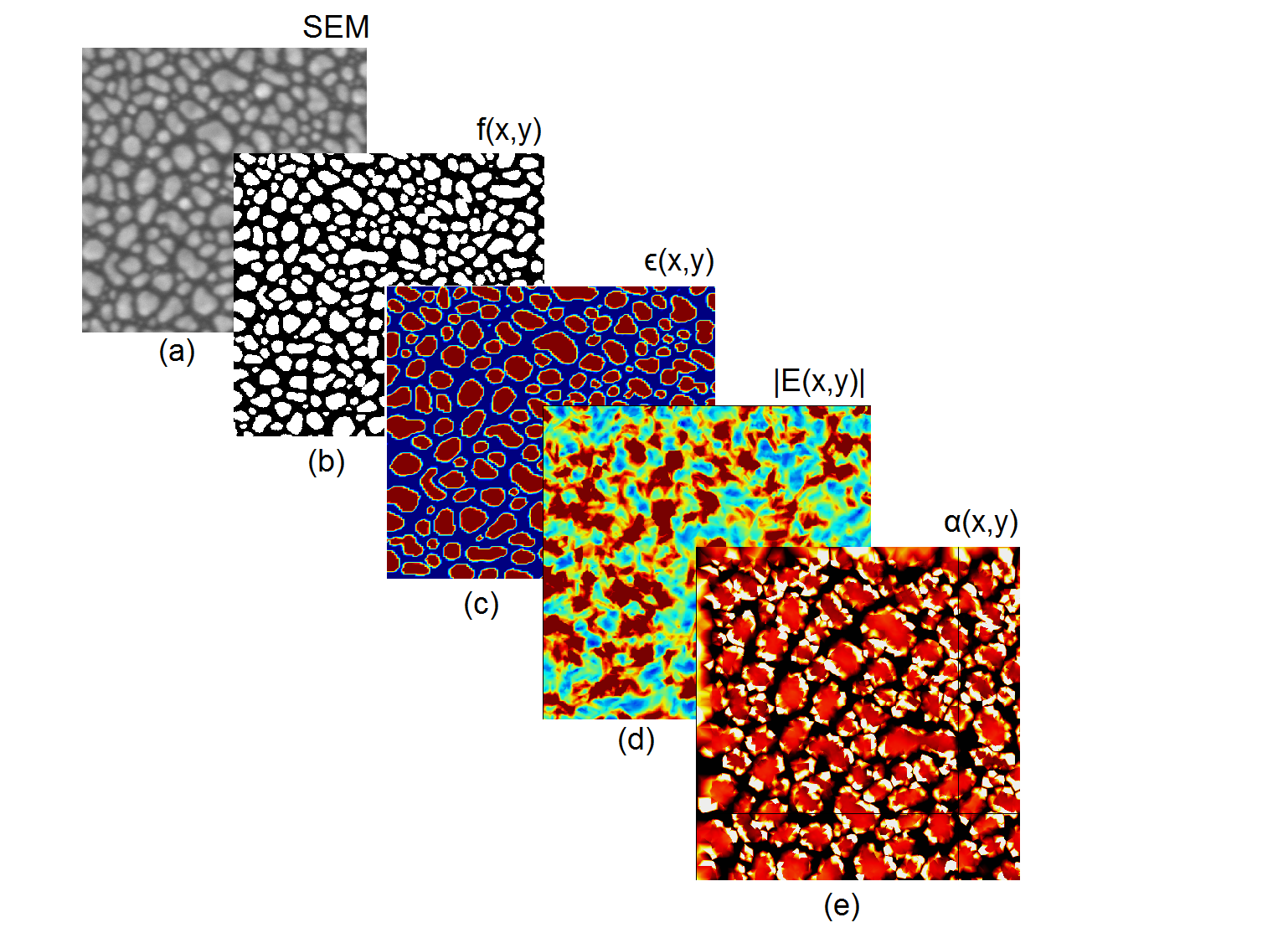}
\caption{A scheme for calculating absorption from SEM image. Convert the SEM (a) into a monochrome bitmap (b), then use this to define a material with ϵ(x,y), a spatially dependant relative permittivity, (c). This material can then be used in FEM calculations to determine the electric field (d) and thus the electromagnetic absorption, $\alpha$, in the layer (e). }
\label{fig:simulationhowto}
\end{figure}

Figure \ref{fig:simulationhowto} shows the scheme used to calculate the total absorption in the metal/semiconductor/metal structures and the absorption of light by the AuNP film and the reflecting layer. 
The first step is to convert the measured SEM image to a monochrome bitmap. 
A function $f$ is then created whose  value  at a particular coordinate $(x,y)$ in the image  is 1 if there is gold at that location and 0 otherwise: 
 \begin{equation}
 f(x,y)=
 \begin{cases}
 0& \text{pixel(x,y) = black},\\
 1& \text{pixel(x,y) = white}.
 \end{cases}
 \end{equation}

With $f$, it is possible to   define a spatially--dependent relative permittivity:
\begin{equation}
\epsilon(x,y,\lambda)= \epsilon_0 + f(x,y)(\epsilon_{Au}(\lambda)-\epsilon_0 ).
\end{equation}

\begin{figure*}[tbph!]
\centering
\includegraphics[width=0.9\linewidth]{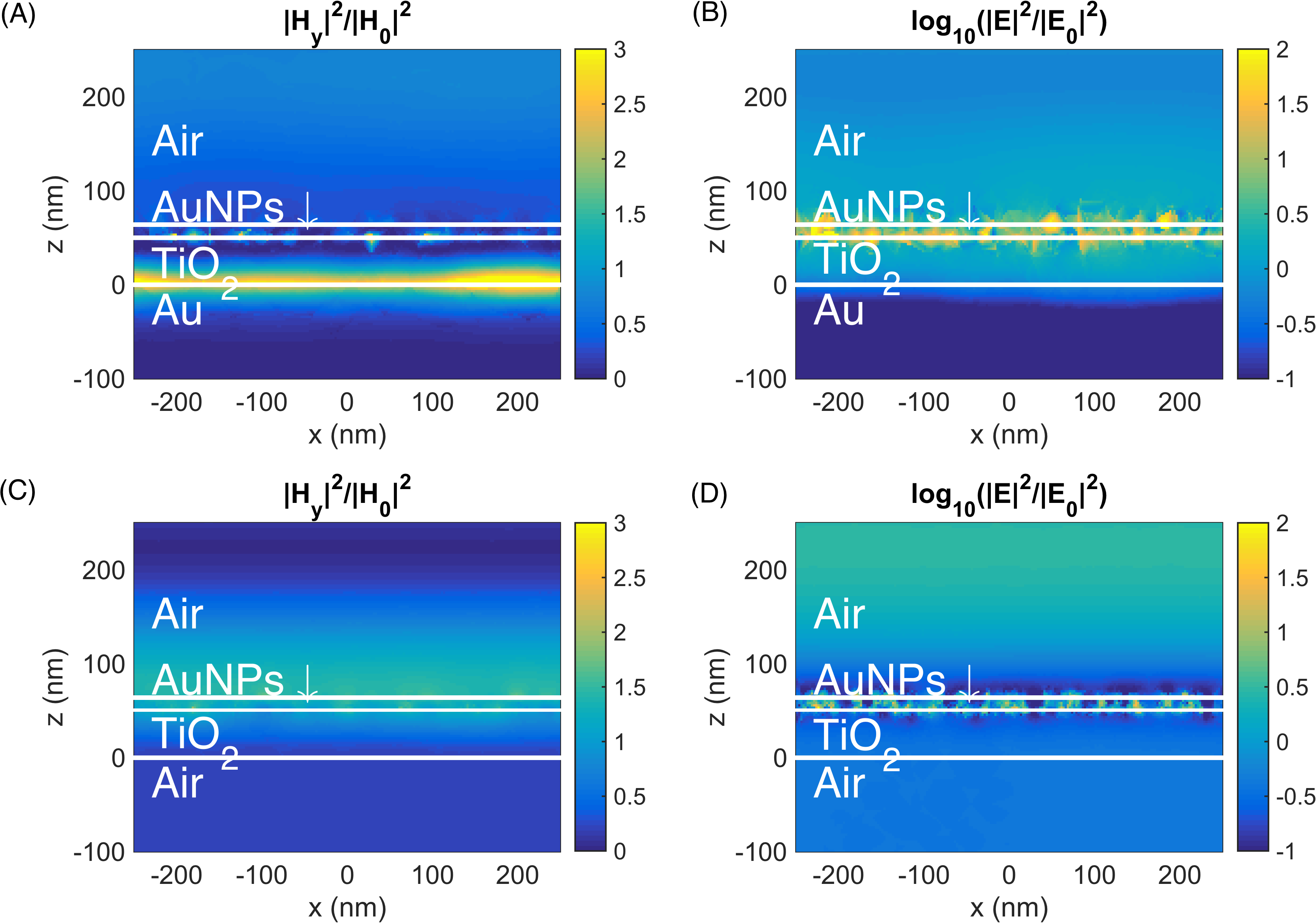}
\caption{spatial distribution  of electromagnetic fields in samples with (A,B) and without (C,D) a metal reflecting layer for a wavelength of 600 nm.}
\label{fig:Fields}
\end{figure*}

Figure \ref{fig:Fields} shows the calculated spatial distribution  of electromagnetic fields in samples with and without a metal reflecting layer.

\cleardoublepage
\section{Model of the internal quantum efficiency $\eta$}
\label{sec:S3}
We consider a top illuminated metal/semiconductor/metal--nanoparticle structure, where the top metal layer consists of metal nanoparticles.
This top layer forms a Schottky contact with the semiconductor with  a barrier height $\Phi_{SB}$ (we  ignore Fermi pinning effects or any other effects due to surface imperfections), and we describe in this section the electron flow that originates from the nanoparticles (possible thermally activated charge flow on the opposite direction is not considered).

The  incident photon--to--electron conversion efficiency is described, to a good level of accuracy, by the product of the absorption efficiency $A(\lambda)$ of the top layer and the internal quantum efficiency $\eta(\lambda)$ for detecting a hot charge carrier per each absorbed photon:
\begin{equation}
\text{IPCE}(\lambda) = A(\lambda) \eta(\lambda).
\end{equation}

$\eta$ is  approximated as the following product:
\begin{equation}\label{eq:S1}
\eta = \eta_\text{inj}\times \eta_\text{trpt}\times\eta_\text{injm}\times\eta_{ed},
\end{equation}
where:
\begin{itemize}

\item $\eta_\text{inj}$: the injection efficiency of hot electrons into the semiconductor material,

\item $\eta_\text{trpt}$ the probability that the electron travels through the semiconductor into the second interface,  

\item $\eta_\text{injm}$ the transmission coefficient for electrons travelling from the semiconductor into the metal electrode/reflector and, 

\item $\eta_\text{ed}$ the efficiency of the redox processes that lead to charge injection from species in solution to positively charged Au nanoparticles. 

\end{itemize}

We now consider each of these processes in detail.

\begin{figure}[tbph!]
\centering
\includegraphics[width=0.9\linewidth]{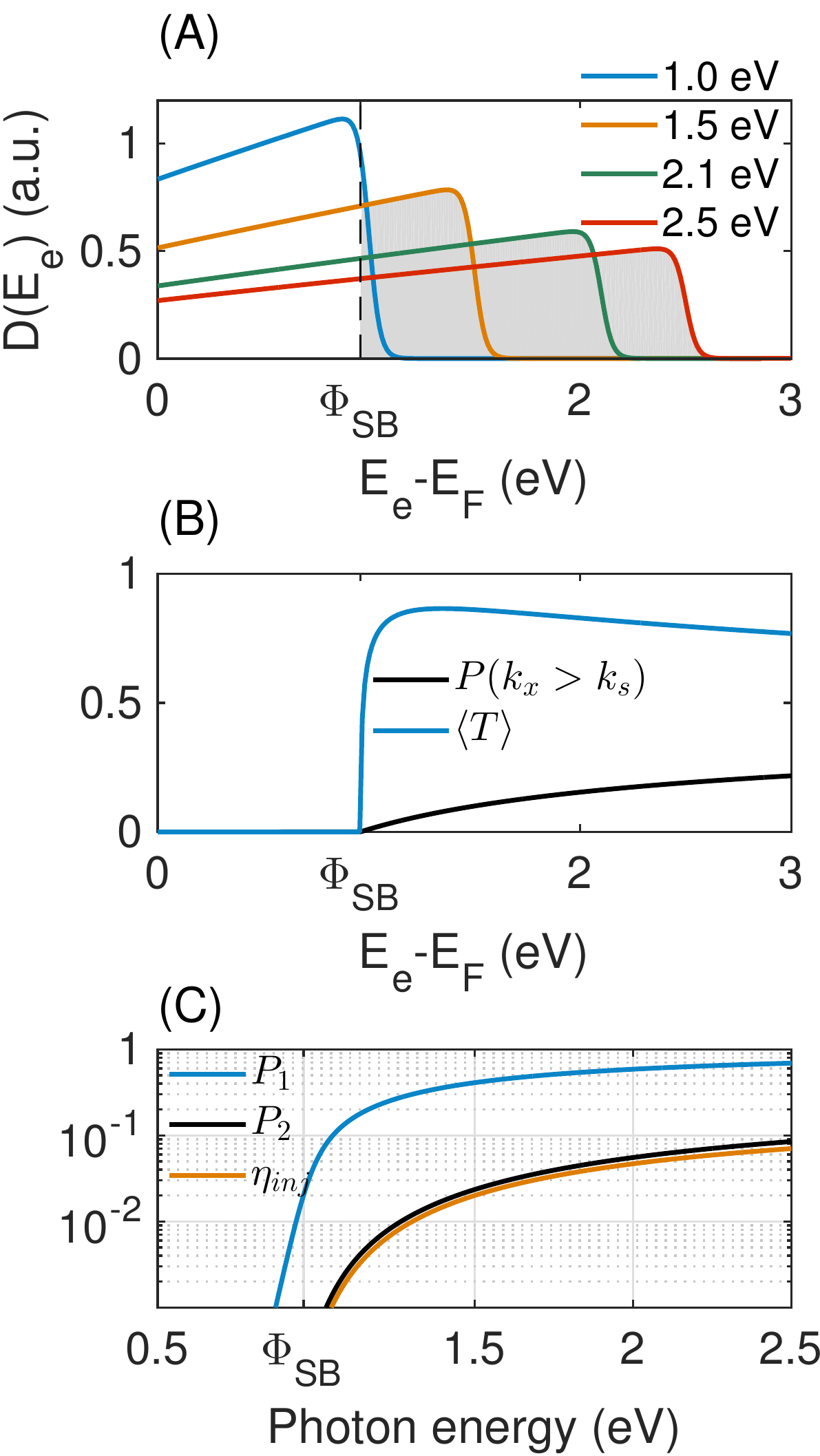}
\caption{Injection efficiency. 
(A) Energy distribution of photo--excited electrons. 
The  horizontal scale represents the excess electron energy with respect to E$_F$. 
The distributions were calculated for different energy values (in eV) of incident photons as indicated in the legend. 
(B) Fraction of the electron population with momenta within  the escape cone ($P(k_x>k_s)$). Transmission coefficient $\langle T\rangle$ shown as a function of excess electron energy.
(C) $P_1$: fraction of the electron population with energies above $\Phi_{SB}$ shown as a function of incident photon energy. 
$P_2$: is the fraction of the electron population with $E_e > \Phi_{SB}$ and with momenta within  the escape cone.
$\eta_{inj}$ is the injection efficiency which corresponds to $P_2$ weighted by the transmission coefficient $\langle T\rangle$.}
\label{fig:STheory1}
\end{figure}

\subsection{$\eta_\text{inj}$: Injection efficiency}

The injection efficiency is the probability of photoexcitation of electrons (by way of surface plasmon relaxation) with  
kinetic energies in excess of the metal--semiconductor Schottky barrier and with sufficient momentum to traverse the barrier.
Injection requires a non--vanishing transmission coefficient across the barrier.

\subsubsection{Energy distribution of photo--excited electrons}

Illumination of the nanoparticles results in the excitation of surface plasmons. 
Non--radiative (Landau) dephasing of these excitations leads to  the energy transfer from a surface plasmon (with energy $h\nu$) to a single electron hole--pair  resulting in the excitation of electrons from below the Fermi level of the metal $E_F$ with energies $E_F - h\nu<E_i<E_F$ to unoccupied states with energies $E_F+E_e$ with $E_e$ being the excess kinetic energy of the electron where $0<E_e< h\nu$.
The energy distribution of these excited states has been argued to be almost uniform due to the fact that Landau damping occurs as intraband transitions between states of $sp$ character that have constant energy--densities a few eV above and below the $E_F$ for metals \cite{Moskovits_RMP1985a}.

Here, we take the approach described by White and Catchpole \cite{White_APL2012a} to estimate the shape of the hot--electron energy density distribution $D(E_e;h\nu)$:
\begin{equation}\label{eq:D(E)}
D(E_e;h\nu) \propto \rho(E_e-h\nu)f(E_e-h\nu)\rho(E_e)\left[1-f(E_e)\right],
\end{equation}
where $\rho(x)$ is well approximated by the free--electron gas model, $\rho(x) \sim x^{1/2}$ and $f(x)$ is the Fermi--Dirac distribution function.
This approximation does not take into account the conservation of momentum in the electronic transitions.
The shape of this distribution is shown in figure \ref{fig:STheory1}(A) for four values of incident photon energy, assuming a metal work function of 5.1 eV and a Schottky barrier height of 1 eV.
The shaded area in these curves correspond to the fraction of the resulting population with energies above the Schottky barrier, a fraction shown \textit{vs} $h\nu$ by curve $P_1$ in figure \ref{fig:STheory1}(C).

\subsubsection{Escape cone in momentum space}
%
%
Let us consider  the metal--semiconductor interface as consisting of an infinite plane in space. 
If we represent with $x$ the axis normal to the interface, then a requirement for electron injection is that the $x$ component of the electron momentum 
$k_{sx} = k_e\cos(\Omega_s)$ 
should have an associated energy larger than the Schottky barrier.
This condition allows us to define an escape cone of angle $\Omega_s$ in $k$ space by:
\begin{equation}\label{eq:escape_cone}
k_{sx} = k_e\cos(\Omega_s) = \sqrt{2m\Phi_{SB}}/\hbar.
\end{equation}

The  fraction of the hot--electron population that posses $x$ components of their momentum with an associated kinetic energy with sufficient magnitude to cross the interface,  is given by the ratio of the solid angle subtended by $\Omega_s$ to the solid angle of the entire sphere of constant energy in $k$ space  (under the assumption of uniform distribution of momenta):
\begin{equation}\label{eq:S_momentum}
\begin{split}
P(k_x>k_s) &= \frac{1}{4\pi}\int_0^{2\pi}\int_0^{\Omega_s}  \sin(\theta)d\theta d\phi\\
&= \frac{1}{2}\left[1-\cos(\Omega_s)\right]\\
&= \frac{1}{2}\left(1-\sqrt{\frac{\Phi_{SB}}{E_e}}\right),
\end{split}
\end{equation}
valid for $E_e>\Phi_{SB}$.
This result shows that the fraction of hot--electrons moving to the flat metal--semiconductor interface asymptotes to 1/2 for the cases where $\Phi_{SB}\rightarrow 0$ or $E_e\gg\Phi_{SB}$.
Figure \ref{fig:STheory1}(B) shows how this fraction of the hot--electron population varies  with electron energy, whereas  figure \ref{fig:STheory1}(C) shows the fraction of the population that meets  both the energy and momentum  requirements for electron injection (curve $P_2$). 

When $D(E_e;h\nu)$ is an uniform distribution, it can be shown that the fraction of the hot--electron population that has both the energy and momentum required for electron escape:
\begin{equation}
\int_{\Phi_{SB}}^\infty D(E;h\nu)P(k_x>k_s),
\nonumber
\end{equation}
asymptotes to the Fowler equation \cite{Scales_QEIJO2010a}.

\subsubsection{Transmission coefficient}
\label{sec:S_T}
One more consideration to bear in mind is the possibility of reflections at the metal--semiconductor interface, which arises due to the possible mismatch between the momenta of the hot--electron in both media.
On the metal side, the kinetic energy of the hot--electron is:
\begin{equation}
E_k =  E_e = \frac{\hbar^2}{2 m_e}\left[k^2\right],
\end{equation} 
whereas on the semiconductor side it becomes
\begin{equation}
E_e - \Phi_{SB} = \frac{\hbar^2 }{2 m_e^*}(\kappa^2),
\end{equation}
where in general $m_e \ne m_e^*$.
The transmission coefficient $T$ at the interface is given by: \cite{Davies_1998a} 
\begin{equation}\label{eq:T}
\begin{split} 
T &= \frac{4k_{sx} \kappa_{x}}{m_e m_e^*(k_{sx}/m_e + \kappa_{x}/m_e^*)^2}\\
&=\frac{4k_{sx} \kappa\cos(\theta)}{m_e m_e^*(k_{sx}/m_e + \kappa\cos(\theta)/m_e^*)^2},
\end{split}
\end{equation}
where $k_{sx}$ is given by Eqn. \eqref{eq:escape_cone} and we have considered  only those hot--electrons within the escape cone  subtended by $\Omega_s$.
$T$ has been written in terms of the total momentum $\kappa$ inside the semiconductor and the angle $\theta$ of its projection on the axis perpendicular to the interface (i.e. $\kappa_x = \kappa\cos(\theta)$).

In the work of Chalabi {\it et al}\cite{Chalabi_NL2014a} it was assumed that due to the translation invariance of their metal--semiconductor interface along one direction, the momentum component parallel to the interface $k_y$  was conserved during the charge transfer process.
This  condition simplifies the calculation of the transmission probability of eqn. \eqref{eq:T}, but it is a condition that may not be satisfied in general (e.g. for metal nanoparticles).
 A more general consideration consists on assuming conservation of the total momentum and accounting for possible changes in the direction of the momentum after injection: the electrons move away from the metal, or equivalently that the angle $\theta$ on eqn. \eqref{eq:T} varies from 0 to $\pi/2$.
 With this in mind, an angle--averaged transmission coefficient $\langle T\rangle$ can be calculated as follows:
 \begin{equation}\label{eq:S_T}
 \begin{split}
 \langle T\rangle &= \frac{1}{2\pi}\int_0^{2\pi}\int_0^{\pi/2}
 \frac{4\alpha\cos(\theta)}{(\alpha + \cos(\theta))^2} d\phi d\theta\\
 &= 4\alpha\left[ \ln\left(\frac{\alpha+1}{\alpha}\right) + \frac{\alpha}{1+\alpha} -1 \right],
 \end{split}
 \end{equation}
where $\alpha = k_{sx} m_e^*/\kappa m_e = \sqrt{\frac{m_e^*\phi_b}{m_e(E_e-\Phi_{SB})}}$.
This transmission coefficient is shown in figure \ref{fig:STheory1}(B) $\langle T\rangle$ assuming an Au--TiO$_2$ interface and a ratio for electron  effective mass of 0.0862 (data taken from Zhang {\it et al }\cite{Zhang_PCCP2014a}).
Similar results are reported by Nienhaus \textit{et al} \cite{Nienhaus_SS2002a}.

The efficiency of hot--electron injection $\eta_\text{inj}$ is then calculated as:
\begin{equation}\label{eq:S_eta_inj}
\eta_\text{inj}(h\nu)  = \frac{\int\limits_{\phi_b}^{\infty}dE D(E;h\nu) P(k_x>k_s)   \langle T(E)\rangle}{\int\limits_0^{\infty}dE D(E;h\nu)}.
\end{equation}

$\eta_\text{inj}$ depends on 
the height of the Schottky barrier height $\Phi_{SB}$, 
the incident photon energy $h\nu$, 
the ratio of electron effective masses in the metal and semiconductor material and 
the temperature.
Figure \ref{fig:STheory1}(C) shows $\eta_\text{inj}$.

\begin{figure}[tbph!]
\centering
\includegraphics[width=0.9\linewidth]{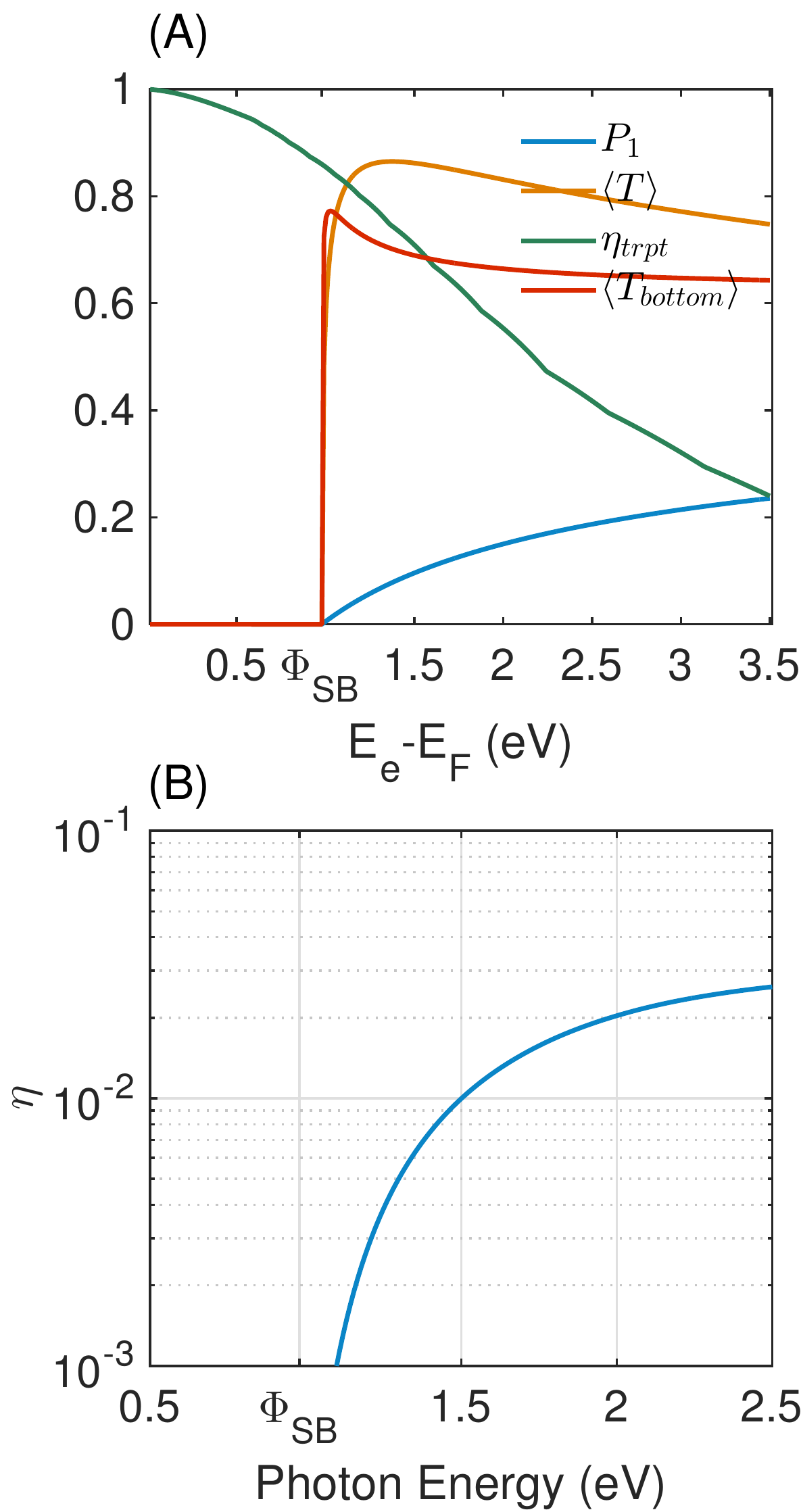}
\caption{Internal quantum efficiency. 
(A) Plots of $\eta_\text{trpt}$ and $\eta_\text{injm}$ shown as a function of electron excess energy above the metal's Fermi level ($E_F$, in eV). For comparison, we also reproduce the curves $P_1$ and $\langle T\rangle$ of figure \ref{fig:STheory2}.
(B) $\eta$ vs incident photon energy for $\Phi_{SB}$=0.98 eV}
\label{fig:STheory2}
\end{figure}

\subsection{$\eta_\text{trpt}$: Transport efficiency}
\label{sec:eta_trpt}
This is the probability that an electron entering the semiconductor layer (of thickness $t$) traverses it without experiencing a scattering event (such as trapping at defect sites).
This depends on the mean free path of charge carriers on the semiconductor material $l_{eff}(E_e)$, which, in turn, depends on the excess energy of the electron $E_e$:
\begin{equation}
\eta_\text{trpt}(E_e) = \frac{1}{t}\int_0^t\exp(-z/l_{eff}(E_e))dz.
\end{equation}
The mean free path for electrons  in TiO$_2$ was obtained at each electron energy from the universal curve compiled in reference \cite{Seah_SAIA1979a}.
For a thickness of 50 nm, the transport efficiency is shown in figure \ref{fig:STheory2}(A), which clearly shows that high energy electrons experience more scattering.

\subsection{$\eta_\text{injm}$: Injection into metal reflector/mirror}
\label{sec:S_eta_injm}

This last step is dictated by   the electron transmission probability at this final interface $\langle T_\text{bottom}\rangle$, which can be calculated in an analogous manner to what was described previously.
Following this rationale, the angle--averaged transmission coefficient is given by:
\begin{equation}
\begin{split}
\eta_\text{injm} &= \langle T_\text{bottom}\rangle \\
&= \frac{(2\beta^2 + 2)\ln(\beta+1)-2\beta^2\ln(\beta)-2\beta}{\beta},
\end{split}
\end{equation}
where:
\begin{equation}
\beta^2 = \frac{m(E_e-\Phi_{SB})}{m^*(E_e-\Phi_{SB}-eV)}.
\end{equation}
where $eV$ is the magnitude of the applied bias across the MSM junction.
The magnitude of this transmission coefficient is shown in figure \ref{fig:STheory2}(A) ($\langle T_{bottom}\rangle$).

\begin{figure}[tbph!]
\centering
\includegraphics[width=0.9\linewidth]{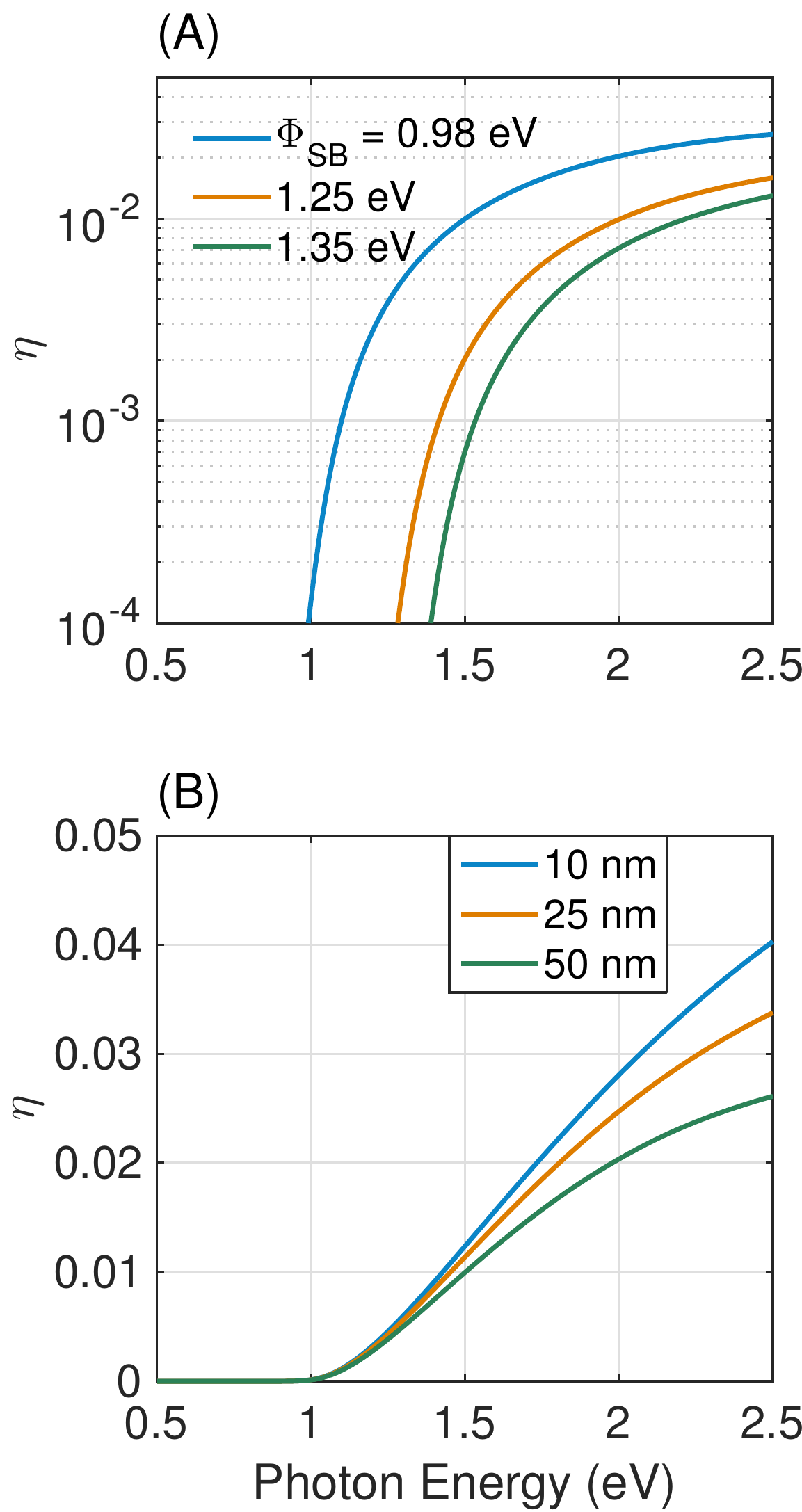}
\caption{Effect of Schottky barrier height (A) and semiconductor thickness (B) on $\eta$}
\label{fig:Stheory3}
\end{figure}

The internal quantum efficiency $\eta$ is then calculated, as a function of incident photon energy as:
\begin{equation*}\label{eq:S_eta}
\eta(h\nu) = \frac{\int\limits_{\phi_b}^{\infty}D(x;h\nu)P(k_x>k_s)\langle T(x)\rangle\eta_\text{trpt}(x)\langle T_{bottom}(x)\rangle dx}{\int\limits_0^{\infty}dE D(x;h\nu)}.
\end{equation*} 
$\eta$ is shown in figure \ref{fig:STheory2}(B).
Figure \ref{fig:Stheory3} shows the expected effect of $\Phi_{SB}$ and semiconductor thickness on $\eta$. 
Similar lineshapes have been obtained by Leenheer {\it et. al}\cite{Leenheer_JOAP2014a}

This model is expected to largely overestimate the IPCE at shorter wavelengths due to the fact that no attempt was made to consider possible photocurrent contributions due to the excitation of surface plasmon polaritons at the Au mirror/TiO$_2$ interface.
As discussed in the main text and by Chalabi \textit{et al}\cite{Chalabi_NL2014a} and Wang and Melosh \cite{Wang_NL2011a}, these currents will have an opposite sign to the ones produced by photoexcitation and plasmon decay  by the metal nanoparticles.


\cleardoublepage


\end{document}